\pdfoutput=1

\documentclass[twocolumn,showpacs,preprintnumbers,pra]{revtex4}
\usepackage{amsmath,amsfonts,amssymb,braket,graphicx}
\usepackage{setspace}
\usepackage[english]{babel}
\usepackage{dcolumn}

\newcommand{\cut}[1]{}
 \newcommand{\kpsi}{\vert \psi\rangle}
 \newcommand{\kxi}{\vert \xi\rangle}
 
 \newcommand{\cH}{\mathcal{H}}
 \newcommand{\cK}{\mathcal{K}} 
 
  \newcommand{\bP}{\mathbf{P}}
 \newcommand{\bS}{\mathbf{S}}
 \newcommand{\bA}{\mathbf{A}}
 \newcommand{\bx}{\mathbf{x}}
 \newcommand{\bAA}{\mathbf{A1}}
 \newcommand{\bAAA}{\mathbf{A2}}

 \newcommand{\Ex}{ \textrm{Ex}}
 \newcommand{\Tr}{ \textrm{Tr}}

  \newcommand{\be}{\beta}
  
  \newcommand{\da}{\dagger}

  \newcommand{\ep}{\epsilon}
  \newcommand{\et}{\eta}

  \newcommand{\rh}{\rho}
  \newcommand{\si}{\sigma}

  \newcommand{\Eq}[1]{Eq.\,(\ref{eq:#1})}
  \newcommand{\eq}[1]{(\ref{eq:#1})}

\def\be {\begin{equation}}
\def\ee {\end{equation}}
\def\1{\mathchoice{\rm 1\mskip-4.2mu l}{\rm 1\mskip-4.2mu l}{\rm
        1\mskip-4.6mu l}{\rm 1\mskip-5.2mu l}}

\def\ket#1{\vert#1\rangle}
\def\bra#1{\langle #1\vert}
\def\braket#1#2{\langle #1\vert#2\rangle}
\newcommand{\ketbra}[1]{\vert#1\rangle\langle#1\vert}

\begin{document}

\title{Violation of Heisenberg's error-disturbance uncertainty relation in neutron spin measurements}
 \author{Georg Sulyok$^1$}
 \author{Stephan Sponar$^1$}
 \author{Jacqueline Erhart$^1$}
\author{Gerald Badurek$^1$}
\author{Masanao Ozawa$^2$}
\author{Yuji Hasegawa$^1$}

\affiliation{%
$^1$Institute of Atomic and Subatomic Physics, Vienna University of Technology, 1020 Vienna, Austria \\
$^2$Graduate School of Information Science, Nagoya University, Chikusa-ku, Nagoya, Japan}

\date{\today}

\begin{abstract}
In its original formulation, Heisenberg's uncertainty principle dealt with the relationship between the error of a quantum measurement and the thereby induced disturbance on the measured object. 
Meanwhile, Heisenberg's heuristic arguments have turned out to be correct only for special cases. A new universally valid relation was derived by Ozawa in 2003. Here, we demonstrate that Ozawa's predictions hold for projective neutron-spin measurements. The experimental inaccessibility of error and disturbance claimed elsewhere has been overcome using a tomographic method. 
By a systematic variation of experimental parameters in the entire configuration space, the physical behavior of error and disturbance for projective spin-$\frac{1}{2}$ measurements is illustrated comprehensively. The violation of Heisenberg's original relation, as well as, the validity of Ozawa's relation become manifest. In addition, our results conclude that the widespread assumption of a reciprocal relation between error and disturbance is not valid in general.
\end{abstract}

%%%%% PACS NUMBERS %%%%%

\maketitle

\section{Introduction}
\label{sec:Introduction}

The uncertainty principle,  proposed by Heisenberg \cite{Heisenberg27} in 1927,  
ranks without doubt among the most famous statements of modern physics. 
The content of the principle is often explained by simply saying 
``The more precisely the position is determined, the less precisely the momentum 
is known, and conversely \cite{Heisenberg27}.''  
It is usually understood that this leads to the impossibility of simultaneous measurements of the position and momentum of a particle
%Since in classical physics the exact knowledge of initial position and momentum is necessary to solve the equations of motions, the uncertainty principle led 
and consequently to the rejection of determinism of the Newtonian mechanics in determining the future motion from the past state.
However, the quantitative formulation of the uncertainty principle has been a target of debate for a long time
(\cite{Arthurs_Kelly_1965, Ballentine_RevModPhys_1970, Busch_JournalTheoreticalPhysics_1985, Kraus_PhysRevD_1987, hilgevoor_uffink, Martens_1990, Martens_1992, Appleby_1998_issue5, Ozawa_PLA_2002} and references therein).

Heisenberg's original formulation \cite{Heisenberg27,Heisenberg30} can be read as the relation
\begin{equation}
\label{eq:Hei27}
\epsilon(Q)\eta(P)\ge\frac{\hbar}{2}
\end{equation}
for the root-mean-square error $\epsilon(Q)$ of a measurement of the
position observable $Q$ and the root-mean-square disturbance $\eta(P)$ of the momentum observable $P$ induced by the position measurement.
However,  modern textbooks usually explain the ``uncertainty principle'' as the relation 
\begin{equation}
\label{eq:Ken27}
\sigma(Q)\sigma(P)\ge \frac{\hbar}{2},
\end{equation}
originally proved by Kennard \cite{Kennard1927} in 1927,
for the standard deviations $\sigma(Q)$ and $\sigma(P)$ of the position observable $Q$ and the momentum observable $P$ in an arbitrary state $\psi$ 
where the standard deviation is defined by 
$\sigma(A)^2=\braket{\psi}{A^{2}|\psi}-\braket{\psi}{A|\psi}^{2}$
for an observable $A$ and a state $\psi$.
Experimental investigations of the above relation can be found in \cite{Shull_PhysRev_1969, Kaiser_PRL_1983, Klein_PRL_1983, Nairz_uncertainty}. 

Heisenberg actually derived Kennard's relation \Eq{Ken27} for Gaussian wave functions $\psi$, applied this relation to the state just after the $Q$ measurement 
with error $\ep(Q)$ and disturbance $\et(P)$, and concluded relation \Eq{Hei27}
from the additional, implicit assumptions $\ep(Q)\ge\si(Q)$ and $\et(P)\ge\si(P)$.
However, his assumption $\ep(Q)\ge\si(Q)$ holds only for a restricted class of measurements \cite{Ozawa_pla_03}.
The assumption $\et(P)\ge\si(P)$ holds if the initial state is the momentum
eigenstate state \cite{Wiseman_foundations1998}, but it does not hold generally.
Thus, his argument did not establish the universal validity of \Eq{Hei27}.

In 1929, Robertson \cite{Robertson29} extended Kennard's relation \Eq{Ken27}
to arbitrary pair of observables $A$ and $B$ as
\begin{equation}
\label{eq:robertson}
\sigma(A)\sigma(B)\ge \frac{1}{2}|\bra{\psi}[A , B]\ket{\psi}|.
\end{equation}

The generalized form of Heisenberg's original relation \Eq{Hei27} would read
\begin{equation}
\label{eq:heisenberg_general}
\epsilon(A) \eta(B) \ge \frac{1}{2}|\bra{\psi}[A ,B]\ket{\psi}|.
\end{equation}
The validity of this relation is known to be limited to specific circumstances \cite{Arthurs_Kelly_1965, Arthurs_PRL_1988, Ishikawa_RepMathPhys_1991, Ozawa_LectureNotes_1991}. In 2003, Ozawa \cite{Ozawa_pra_03, Ozawa_AnnPhys_04, Ozawa_pla_03, Ozawa_jso_05}, one of the authors of the present paper, thus proposed a new error-disturbance uncertainty relation
\begin{equation}
    \label{eq:ozawa}
    \epsilon(A)\eta(B)+\epsilon(A)\sigma(B)+\sigma(A)\eta(B)\ge \frac{1}{2}|\bra{\psi}[ A , B]\ket{\psi}|
\end{equation}
and proved the universal validity in the general theory of quantum measurements,
where $\epsilon(A)$ is the root-mean-square error of an arbitrary measurement for an observable $A$, $\eta(B)$ is the root-mean-square disturbance on another observable $B$ induced by the measurement, $\si(A)$ and $\si(B)$ stands for the standard deviations of $A$ and $B$ in the state $\psi$ just before the measurement.

There appeared questions on the accessibility of errors and disturbances in practical experiments, e.g. \cite{Werner2004,Koshino2005}. This difficulty has been overcome in two ways: the ``three state method'' and the ``weak measurement method.'' The tomographic ``three state method'' is based on a formula to statistically determine error and disturbance, respectively, in a given initial state from the experimentally accessible data from three different auxiliary input states generated from the initial state \cite{Ozawa_AnnPhys_04}.
The ``three state method'' has already been successfully implemented in our previous publication  \cite{Erhart12}. The ``weak measurement method'' is proposed \cite{lund_wiseman_njop2010} to quantify the error and disturbance exploiting the weak-measurement technique \cite{Mir_NJoP_2007} based on the relation between mean-square differences and weak joint distributions \cite{Ozawa_PLA_2005};
an optical experiment was carried out recently \cite{Steinberg_PRL_2012}.

In this paper, we report a general test of the validity of Ozawa's formulation in the framework of projective spin-$\frac{1}{2}$ measurements. Polarized neutrons experience successive spin-measurements A and B, where the former measurement is detuned on purpose thus causing the measurement error of the A measurement. Likewise, the disturbance on the observable $B$ also changes when the A-measurement is detuned. In \cite{Erhart12}, we focussed solely on a small subset of the possible configurations in parameter space where Heisenberg's original relation \Eq{heisenberg_general} is always violated and the typical reciprocal trade-off between the error and the disturbance is maintained.
Here, we give a more complete view on how systematic variations of the observables, the settings of the measurement apparatus, and the pre-measurement system state influence error, disturbance, and pre-measurement standard deviations. By studying the whole configuration space the physical meaning of error and disturbance for projective spin-$\frac{1}{2}$ measurements can be illustrated quite comprehensively.  Parameter regions where Heisenberg's original formulation is violated as well as the behaviour of Ozawa's relation become manifest: not only the validity but also the analytic dependence of two notions of inequalities, \Eq{heisenberg_general} and \eq{ozawa}, on the experimental parameters is clearly seen. In addition, investigating error and disturbance in the whole configuration space shows the restricted validity of a reciprocal relation between these quantities in the spin-$\frac{1}{2}$ system.

\section{Theory}
\label{sec:theory}

In order to derive the universally valid error-disturbance uncertainty relation 
it is convenient to discuss general models of measuring processes described 
in the quantum mechanical framework.  For the detailed account, we refer the
reader to \cite{Ozawa_AnnPhys_04, Ozawa_pra_03, Ozawa_pla_03, Ozawa_jso_05}. 
Here, we will sketch the main results necessary for later discussions on our 
experimental work.

\subsection{Indirect measurement models}
In this paper, we consider only {\em finite-type quantum measurements} 
such that the measured system is described by a finite dimensional Hilbert space 
and that the apparatus has a finite number of possible outcomes.
We assume that every measuring apparatus has its own output variable $\bx$.
The apparatus $\bA(\bx)$ having the output variable $\bx$ is assumed to 
determine the probability distribution $\Pr\{\bx=m\|\rho\}$ of $\bx$
on every input state $\rho$, and to determine the output state $\rho_{\{\bx=m\}}$
for every input state $\rho$ conditional upon each possible output $\bx=m$.

An {\em indirect measurement model} of an apparatus $\bA(\bx)$ measuring
a system $\bS$ described by a Hilbert space $\cH$
is specified by a quadruple $(\cK,\kxi,U,M)$ 
consisting of a Hilbert space $\cK$ describing the probe system $\bP$, 
a state vector $\kxi$ in $\cK$ describing the initial state of $\bP$, 
a unitary operator $U$ on $\cH\otimes\cK$ describing the time
evolution of the composite system $\bS+\bP$ during the measuring interaction,
and an observable $M$, called the {\em meter observable}, 
of $\bP$ describing the meter of the apparatus  \cite{Ozawa_JMathPhys_1984}.
%Note that the meter observable $M$ is to be actually measured 
%after the measuring interaction by an external observer and 
%that measuring process for $M$ no more interacts with $\bS$.

The class of indirect measurement models is a universal class of models of quantum
measurement \cite{Ozawa_JMathPhys_1984, Ozawa_AnnPhys_04} 
in the sense that for any apparatus $\bA(\bx)$ with the output variable $\bx$
there is an indirect measurement model  $(\cK,\kxi,U,M)$ that determines the statistical
properties of  $\bA(\bx)$ by 
\begin{equation}
\label{eq:instrument}
\Pr\{\bx=m\|\rho\}\rho_{\{\bx=m\}}=
\Tr_{\cK}[U^{\dagger}(\1\otimes E^{M}(m))U(\rho\otimes\ketbra{\xi})],
\end{equation}
where $\Tr_{\cK}$ stands for the partial trace over $\cK$ and $E^{M}(m)$
the spectral projection of $M$ corresponding to the real number $m$.
From \Eq{instrument} we have
\begin{eqnarray}
\Pr\{\bx=m\|\rho\}&=&
\Tr[U^{\dagger}(\1\otimes E^{M}(m))U(\rho\otimes\ketbra{\xi})],\\
\rho_{\{\bx=m\}}&=&
\frac{\Tr_{\cK}[U^{\dagger}(\1\otimes E^{M}(m))U(\rho\otimes\ketbra{\xi})]}
{\Tr[U^{\dagger}(\1\otimes E^{M}(m))U(\rho\otimes\ketbra{\xi})]}.\qquad
%. {MO0504}
\end{eqnarray}

\subsection{Measurement operators}
In this paper, we treat only the case where the meter observable $M$
has non-degenerate eigenvalues.  In this case, $M$ has a spectral decomposition
$M=\sum_{m}m\ket{m}\bra{m}$, where $m$ varies over eigenvalues of $M$.
Then, the apparatus $\bA(\bx)$ has a family $\{M_{m}\}$ of operators, called
the {\em measurement operators}, defined by $M_m=\bra{m}U\ket{\xi}$.   
We have
\begin{eqnarray}
U\ket{\psi}\ket{\xi}
&=&\sum_{m}M_{m}\ket{\psi}\ket{m}.
\end{eqnarray}
This means that on an input vector state $\kpsi$ the apparatus $\bA(\bx)$ outputs 
the outcome $\bx=m$ 
with probability $\Pr\{\bx=m\|\kpsi\}=\|M_{m}\kpsi\|^2$, where  
$\| \cdots \|$ denotes the Euclidean norm given by 
$\|\ket{\phi}\|=\langle\phi|\phi\rangle^{\frac{1}{2}}$,
and leaves $\bS$ in the vector state 
\begin{equation}
\ket{\psi}_{\{\bx=m\}}
=\frac{M_{m}\ket{\psi}}{\|M_{m}\ket{\psi}\|}.
\end{equation}
The  {\em probability operator-valued measure (POVM)} of  $\bA(\bx)$
is the family $\{\Pi_{m}\}$ of operators defined by 
$\Pi_{m}=M^{\dagger}_{m}M_{m}$.
Then, we have 
$\Pr\{\bx=m\|\ket{\psi}\}=\|\Pi_{m}^{1/2}\ket{\psi}\|^2=\braket{\psi}{\Pi_m|\psi}$.
The {\em non-selective operation} of $\bA(\bx)$ is a trace-preserving
completely positive map $T$ defined by 
\begin{equation}
\label{eq:non_selective_operation}
T\rh=\sum_{m}M_{m}\rho M_{m}^{\dagger}
\end{equation}
for any state $\rho$; the state $T\rh$ is the state just after the measurement
without post-selection for the input state $\rho$. 

\subsection{Universal uncertainty principle}

Let $A,B$ be observables of $\bS$.
We consider the measurement of the observable $A$ carried out by the 
apparatus $\bA(\bx)$ and the disturbance on $B$ caused by this measurement.
If the input state of $\bS$ is $\ket{\psi}$, 
the {\em root-mean-square (rms) error} $\epsilon(A)$ of $\bA(\bx)$ 
for measuring an observable $A$ of $\bf{S}$ 
and the {\em rms disturbance} $\eta(B)$ of $\bA(\bx)$ 
caused on an observable $B$ of $\bf{S}$ are defined as
\begin{eqnarray}
\epsilon(A)&=&\|\,(U^{\dagger}(\1\otimes M)U-A\otimes \1)\kpsi\kxi\|,
\label{eq:error_def}\\
\eta(B)&=&\|\,(U^{\dagger}(B\otimes \1)U-B\otimes \1)\kpsi\kxi\|,
\label{eq:dist_def}
\end{eqnarray}
where $\1$ stands for the identity operator on the respective space.
The error $\epsilon(A)$ is the root-mean-square of the difference
between the meter observable $M$ after the interaction and
the observable $A$ before the interaction. 
The disturbance $\eta(B)$ is the root-mean-square of the change in the 
observable $B$ during the measuring interaction.
Then, it is mathematically proved \cite{Ozawa_pra_03,Ozawa_AnnPhys_04} that 
\begin{equation}
    \label{OZA}
    \epsilon(A)\eta(B)+\epsilon(A)\sigma(B)+\sigma(A)\eta(B)\ge \frac{1}{2}|\bra{\psi}[ A , B]\ket{\psi}|,
\end{equation}
holds for any state $\kpsi$ of $\bS$ and
any indirect measurement model $(\cK,\kxi,U,M)$.

We define the {\em $k$-th moment output operator}  by
\begin{equation}
O_A^{(k)}=\sum_m m^k \Pi_m=\sum_m m^k M_m^{\da}M_m.
\label{eq:OA} %{KE.0411}
\end{equation}
Then $k$-th moment of the output variable $\bx$ of $\bA(\bx)$ on
input $\kpsi$ is given by 
\begin{equation}
\Ex\{\bx^k\|\kpsi\}=\langle\psi\vert O_A^{(k)}|\psi\rangle
\end{equation}
where "Ex" abbreviates expectation value.
We have \cite{Ozawa_AnnPhys_04}
\begin{eqnarray}\label{eq:e}
\epsilon(A)^2&=&
\bra{\psi}O_A^{(2)}-O_A^{(1)}A-AO_A^{(1)}+A^2\ket{\psi}.
\end{eqnarray}
We define the {\em post-measurement $k$-th moment operator} 
of the observable $B$ by
\begin{eqnarray}
O_B^{(k)}&=&\sum_m M_m^{\dagger}B^{k}M_m.
\label{eq:OB} %{KE.0411}
\end{eqnarray}
Then the $k$-th moment of the observable $B$ in the state 
on output from the apparatus $\bA(\bx)$ on 
input $\kpsi$ is given by 
\begin{equation}
\Ex\{B^k\|T(\ketbra{\psi})
\}=\langle\psi\vert O_B^{(k)}|\psi\rangle,
\end{equation}
where $T$ is the non-selective operation of  $\bA(\bx)$ (see \Eq{non_selective_operation}).
Then, we have \cite{Ozawa_AnnPhys_04}
\begin{eqnarray}\label{eq:d}
\eta(B)^2&=&
\bra{\psi}O_B^{(2)}-O_B^{(1)}B-BO_B^{(1)}+B^2\ket{\psi}.
\end{eqnarray}
Therefore, both error $\ep(A)$ and disturbance $\et(B)$ are determined
by the measurement operators of the apparatus $\bA(\bx)$.

With the help of $\{M_{m}\}$ we can rewrite error and disturbance starting 
from their definitions Eqs.\,\eq{error_def} and \eq{dist_def} to
\begin{eqnarray}
\label{eq:error_Mm}
\epsilon(A)^2&=&\sum_{m} \|M_{m} ({m}-A)\ket{\psi}\|^2
\\
\label{eq:dist_Mm}
\eta(B)^2&=&\sum_{m} \|[M_{m}, B]\ket{\psi}\|^2
\end{eqnarray}
Details of the calculation can be found in Sec.4.4 of \cite{Ozawa_jso_05}.
If the ${M_{m}}$ are mutually orthogonal projection operators 
the sum in \Eq{error_Mm} can by using the Pythagorean theorem be replaced by 
\begin{equation}
\epsilon(A)^2 = \| \sum_{m} M_{m} ({m}-A)\ket{\psi}\|^2
=\|(O_A - A)\ket{\psi}\|^2
\label{eq:error_proj}
\end{equation}
where we used $\sum_{m}M_{m}=\1$ and defined the output operator $O_A =\sum_{m} m M_{m}$ being just the $1^{\rm st}$ moment output operator $O_A^{(1)}$ in case of projective measurements. In analogous manner, we will often abbreviate $O_B^{(1)}$ as $O_B$.

\subsection{Three-state method for quantifying error and disturbance}
The  error $\ep(A)$ and the disturbance $\et(B)$ have been defined through
the noise operator $N(A)=U^{\dagger}(\1\otimes M)U-A\otimes \1$ and 
the disturbance operator $D(B)=U^{\dagger}(B\otimes \1)U-B\otimes \1$, respectively.
However, given an apparatus $\bA(\bx)$, those operators are usually
unknown in practice.
It is also impossible to measure $N(A)$ by measuring $A\otimes \1$ and 
$U^{\dagger}(\1\otimes M)U$ successively, since those two observables 
may not commute.
Similarly, it is also impossible to measure $D(B)$ by successive measurement.
The {\em three-state method for quantifying error and disturbance}
is a method to measure $\ep(A)$ and $\et(B)$ using the outcomes from the
apparatus $\bA(\bx)$ but without knowing either the 
noise operator  $N(A)$ or the disturbance operator $D(B)$.

From \Eq{e} the error $\ep(A)$ can be written as \cite{Ozawa_AnnPhys_04}
\begin{eqnarray}
\label{eq:error_5terms}
\epsilon(A)^{2}&=&\langle \psi |A^2| \psi \rangle + \langle \psi | O_{A}^{(2)} | \psi \rangle + \langle \psi | O_{A}^{(1)} |\psi  \rangle  \nonumber  \\
& &
+\langle A \psi |O_{A}^{(1)} | A\psi \rangle 
 - \langle (A+\1) \psi |O_{A}^{(1)} |(A+\1)\psi\rangle,\nonumber\\
\end{eqnarray}
where we use such abbreviations as $\ket{A\psi}\equiv A\ket{\psi}$, etc. 
Thus, $\ep(A)$ can be statistically estimated from the measurement 
outcomes from the apparatus $\bA(\bx)$ 
in the three states $\kpsi$, $A\kpsi / \Vert A \kpsi \Vert$, 
and $(A+\1)\kpsi/\Vert (A+\1)\kpsi \Vert$ on input.
Similarly, from \Eq{d} the disturbance $\et(B)$ can be written as 
\begin{eqnarray}
\label{eq15}\label{eq:dist_5terms}
\eta(B)^{2}&=& \langle \psi |B^2| \psi \rangle + \langle \psi | O_{B}^{(2)} 
| \psi \rangle+ \langle \psi | O_{B}^{(1)} |\psi  \rangle
\nonumber   \\ 
 & &
  +\langle B \psi |O_{B}^{(1)} | B \psi \rangle 
 - \langle (B+\1) \psi |O_{B}^{(1)} |(B+\1)\psi\rangle.\nonumber\\
\end{eqnarray}
Thus, $\et(B)$ can be statistically estimated from the measurement 
outcomes of the $B$-measurement in the state just after passing through
the the apparatus $\bA(\bx)$ in the three states 
$\kpsi$, $B\kpsi / \Vert B \kpsi \Vert$ 
and $(B+\1)\kpsi/\Vert (B+\1)\kpsi \Vert$ 
on the input of $\bA(\bx)$.
According to the theory of operations \cite{Kraus_Lecture_Notes_1983}, 
for a general observable $X$ the state $X\ket{\psi}$ can be prepared 
from the state $\kpsi$
with success
probability $\|X\ket{\psi}\|^2/\|X\|^2$ 
even without knowing the state $\ket{\psi}$.
In fact, an apparatus $\bA(\bx)$ with measurement operators $\{M_0,M_1\}$
with $M_0=X/\|X\|$ and $M_1=\sqrt{\1-|M_0|^2}$  produces the state 
$X\kpsi/\|X\kpsi\|$ on input state $\kpsi$ with probability 
$\Pr\{\bx=0\|\kpsi\}=\|X\ket{\psi}\|^2/\|X\|^2$.

\section{Theory for spin measurements}
\label{sec:theory_spin}

\begin{figure}[!htb]
\begin{center}
\includegraphics[width=8.5cm,keepaspectratio=true]{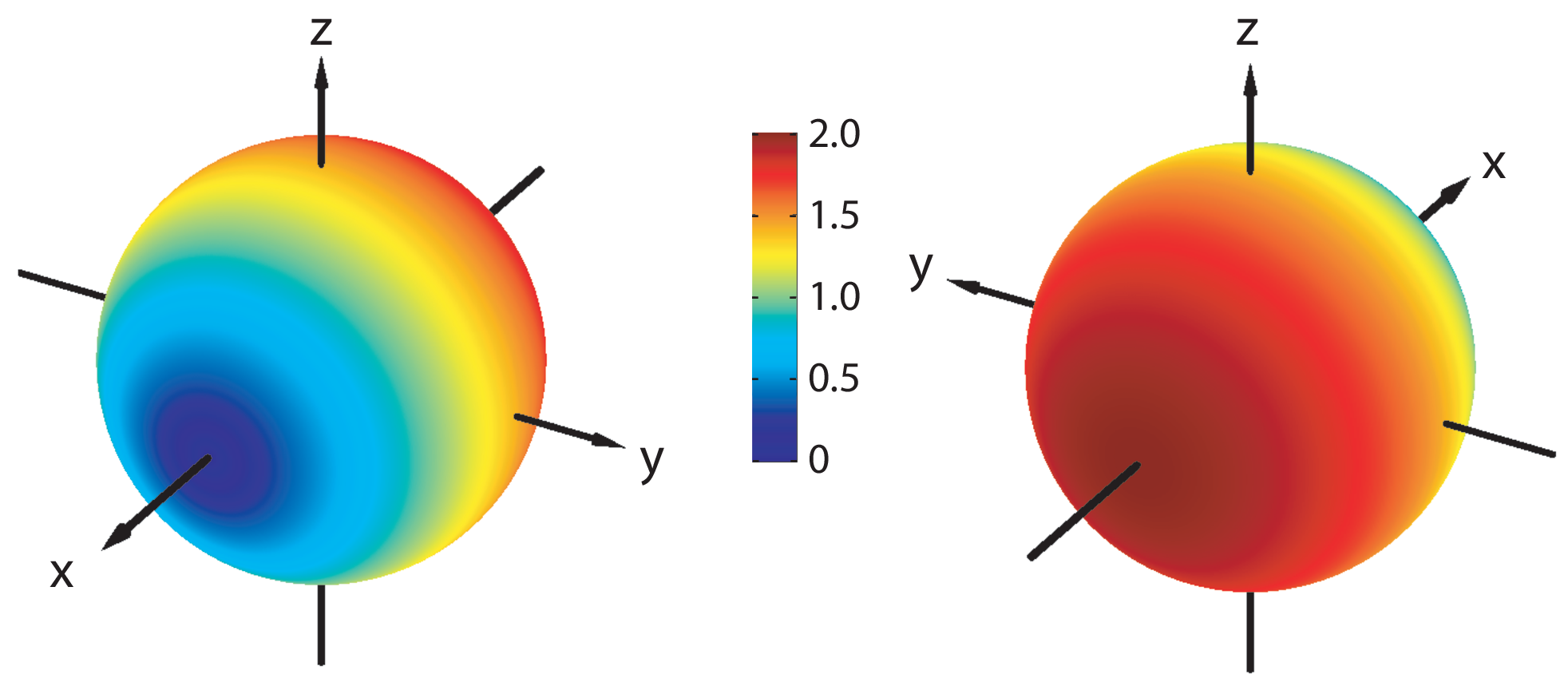}
\caption{(Color online) Error $\epsilon(A)$ of the $A$-measurement: the direction $\vec a$ of $A$ ( $A$ represents the observable to be measured) is fixed to be $(1,0,0)^T$ and the points on the Bloch sphere indicate the direction of $O_A$ (that is the observable actually measured by the apparatus). The value of corresponding error for this combination of vectors $\vec a$ and $\vec o_a$ is color-encoded resulting in concentric circles around the axis defined by $\vec a$.}
\label{fig:error_bloch}
\end{center}
\end{figure}

We will now apply the general results of the previous section to
projective spin-$\frac{1}{2}$ measurements. The observables under consideration are spins along two different directions $\vec a$ and $\vec b$, that is,
\begin{eqnarray}
A=\vec a \cdot \vec\sigma, \qquad B = \vec b \cdot \vec\sigma
\end{eqnarray}
where $ \vec\sigma = (\sigma_x,\sigma_y,\sigma_z)^T$ denotes the vector of the Pauli matrices.
The apparatus is supposed to carry out a projective spin measurement along a distinct axis $\vec o_a$. The output operator $O_A (=O_A^{(1)})$ and the family of measurement operators $\{M_m\}$ are then given explicitly by
\begin{eqnarray}
O_A=\vec o_a \cdot \vec\sigma = M_{+1}-M_{-1},\quad
M_{\pm 1}=\frac{1}{2}(\1 \pm \vec o_a \cdot \vec\sigma)
\end{eqnarray}
Inserting these expressions into Eqs.\,\eq{error_Mm} and \eq{dist_Mm} we get for error and disturbance
\begin{eqnarray}
\label{eq:error_spin}
\epsilon(A)&=&\sqrt{2-2 \ \vec a \cdot \vec o_a}
= 2 \, |\sin\frac{\alpha}{2}|
\\
\label{eq:dist_spin}
\eta(B)&=&\sqrt{2-2 (\vec b \cdot \vec o_a)^2}=
\sqrt 2 \, |\sin\beta|
\end{eqnarray}
where $\alpha$ denotes the angle between $\vec a$ and $\vec o_a$, $\beta$ the angle between $\vec b$ and $\vec o_a$, and $|..|$ the modulus.
Both error and disturbance are thus independent of the initial system state, they are solely determined by the angle between the direction of the observable and the direction of the output operator, that is, the apparatus' measurement axis. The error vanishes if $O_A=A$ and reaches its maximal value ($\epsilon=2$) if $A$ and $O_A$ point in opposite directions. The whole expression can be illustrated on the Bloch sphere (see Fig.\,\ref{fig:error_bloch}).

\begin{figure}[!htb]
\begin{center}
\includegraphics[width=8.5cm,keepaspectratio=true]{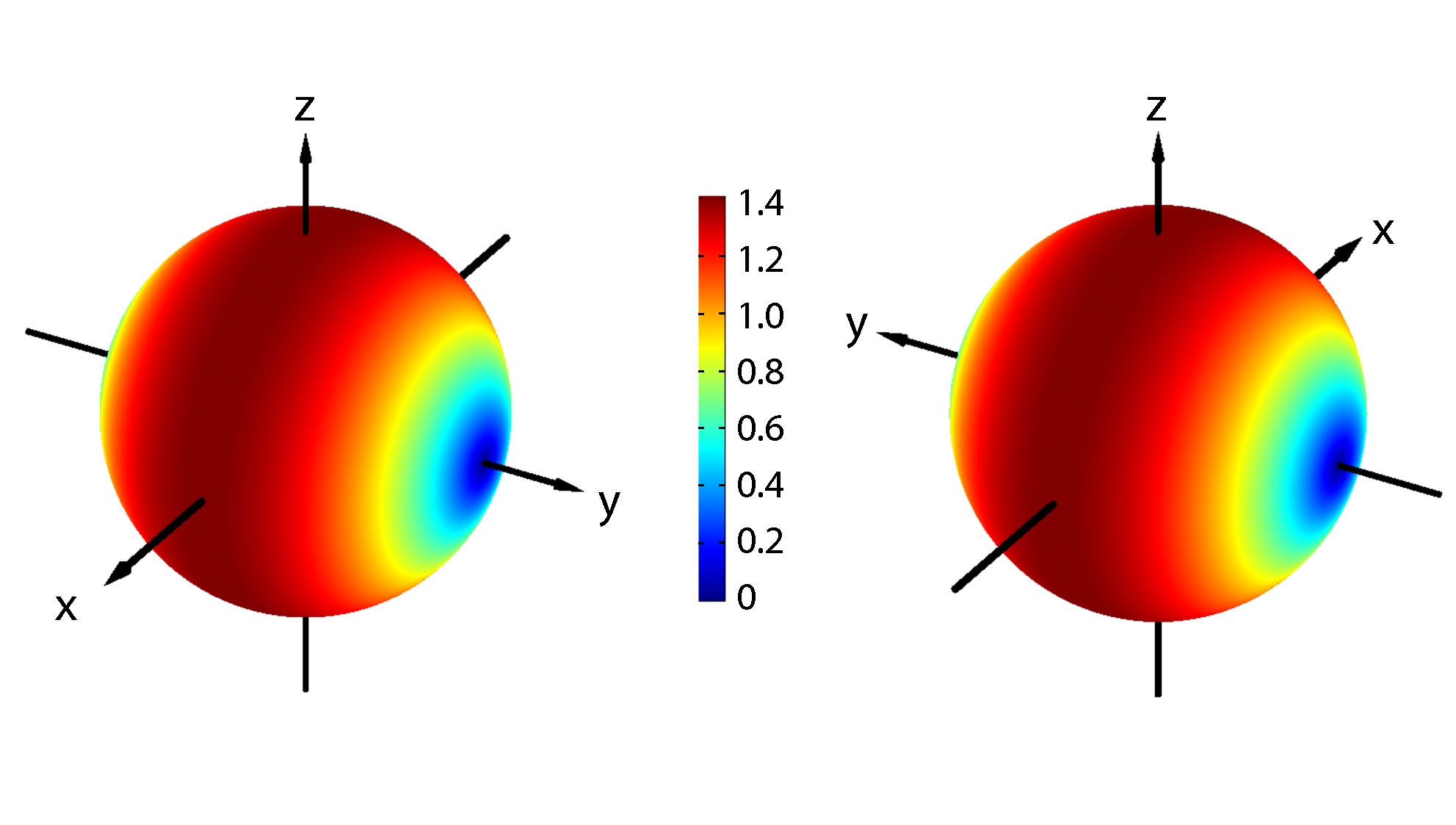}
\caption{(Color online) Disturbance $\eta(B)$ on observable $B=\vec b \cdot \vec\sigma=\sigma_y$ after the projective measurement of $O_A$. Every point on the Bloch sphere corresponds to a possible direction of $O_A$. The colour of this point on the Bloch sphere indicates the value of the disturbance.}
\label{fig:dist_bloch}
\end{center}
\end{figure}

The disturbance induced on $B$ by the prior measurement of $O_A$ is zero if $B$ and $O_A$ point in the same direction, but also if $O_A$ points exactly in the opposite direction, that is if $\vec o_a = -\vec b$. In both cases $O_A$ and $B$ have an identical set of eigenvectors leading to a vanishing disturbance (see \Eq{dist_Mm}). That's a notable difference between the physical concepts of error and disturbance. If the two endpoints of $\vec b$ and $-\vec b$ define the poles on the Bloch sphere, the disturbance reaches its maximal value ($\eta=\sqrt 2$) if $O_A$ lies on the corresponding equator (illustrated in Fig.\,\ref{fig:dist_bloch} for $B=\sigma_y$).

\begin{figure}[!htb]
\begin{center}
\includegraphics[width=8.5cm,keepaspectratio=true]{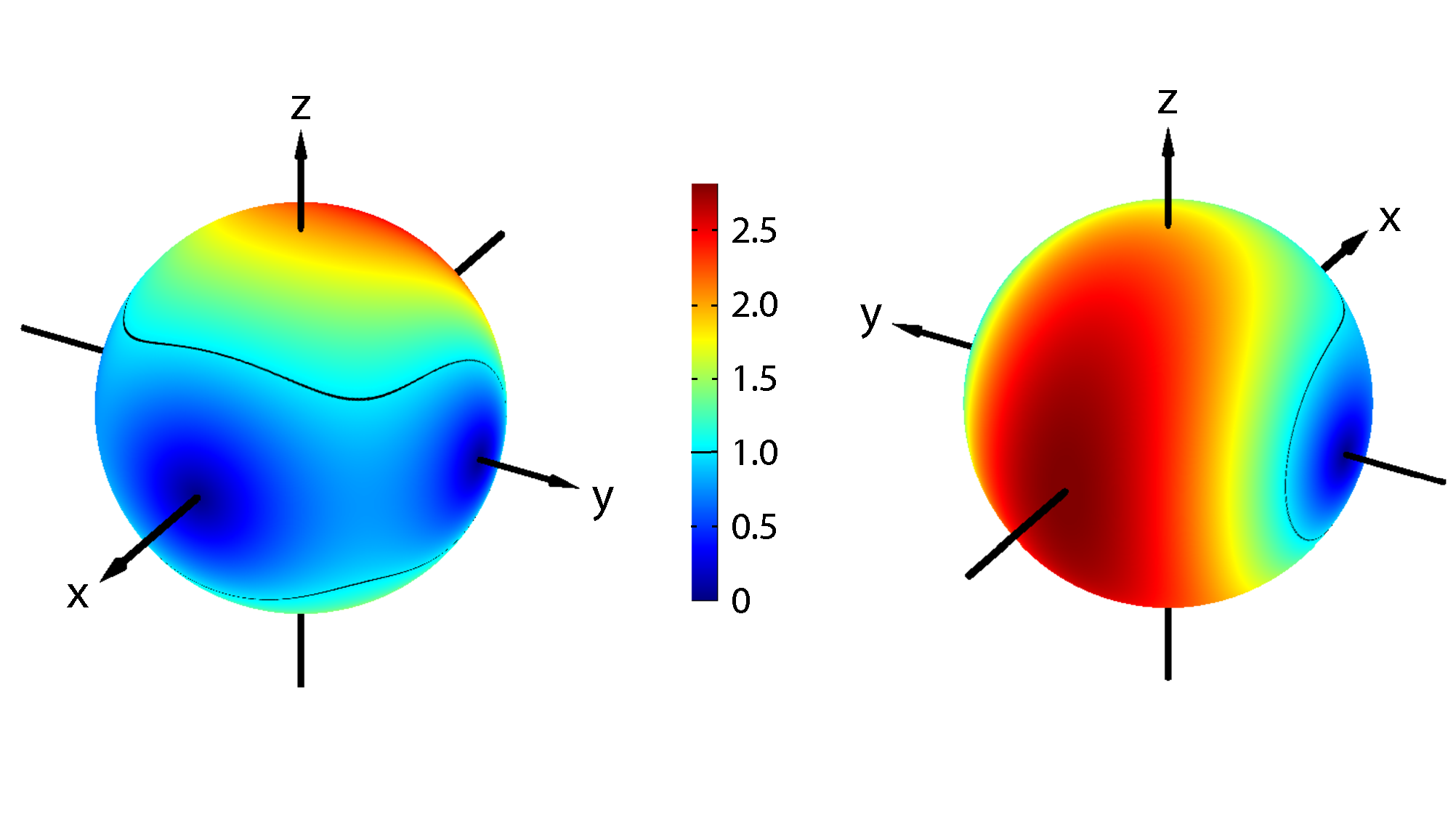}
\caption{(Color online) The product of error and disturbance for $A=\sigma_x$ and $B=\sigma_y$. Every point on the Bloch sphere corresponds to a possible direction of $O_A$ and its color encodes the value of $\epsilon(A)\eta(B)$. The black line represents the maximal lower bound given by the commutator $\frac{1}{2}|\bra{\psi}[A,B]\ket{\psi}|=1$ for $\ket{\psi}=\ket{+z}$. Only if the apparatus is so strongly detuned that it measures  $O_A$  along a a direction outside of the region enclosed by this contour line Heisenberg's error-disturbance relation \Eq{heisenberg_general} is fulfilled.}
\label{fig:errordist_bloch}
\end{center}
\end{figure}

Since error and disturbance are both independent from the initial system state their product $\epsilon(A)\eta(B)$ also depends only on the relative orientation of $\vec a, \vec b$, and $\vec o_a$ and can be depicted on the Bloch sphere for fixed $\vec a$ and $\vec b$ (see Fig.\,\ref{fig:errordist_bloch}). In order to test if a generalized Heisenberg-like relation (\Eq{heisenberg_general}) is obeyed we also have to evaluate the state-dependent limit. By assigning the vector $\vec r$ to the initial spin state using the Bloch representation of its corresponding density matrix $\ket{\psi}\bra{\psi}=\frac{1}{2}(\1 + \vec r\cdot\vec \sigma)$ we get
\begin{equation}
\label{eq:comutator_spins}
\frac{1}{2}|\bra{\psi}[A , B]\ket{\psi}|= |\vec r \cdot (\vec a \times \vec b)|
\end{equation}
The limit is thus given by the volume of the parallelepiped spanned by the three vectors $\vec a$, $\vec b$, and $\vec r$ and ranges from 0 to 1. Therefore, except for the special cases where it vanishes, we can always find regions where the product of error and disturbance is below the limit. This is an example for a general result (chap.6 in \cite{Ozawa_jso_05}) stating that projective measurements of $A$ violate a Heisenberg-like error-disturbance uncertainty relation provided that $B$ is bounded and $\bra{\psi}[A,B]\ket{\psi} \neq 0$.

\begin{figure}[!htb]
\begin{center}
\includegraphics[width=8.5cm,keepaspectratio=true]{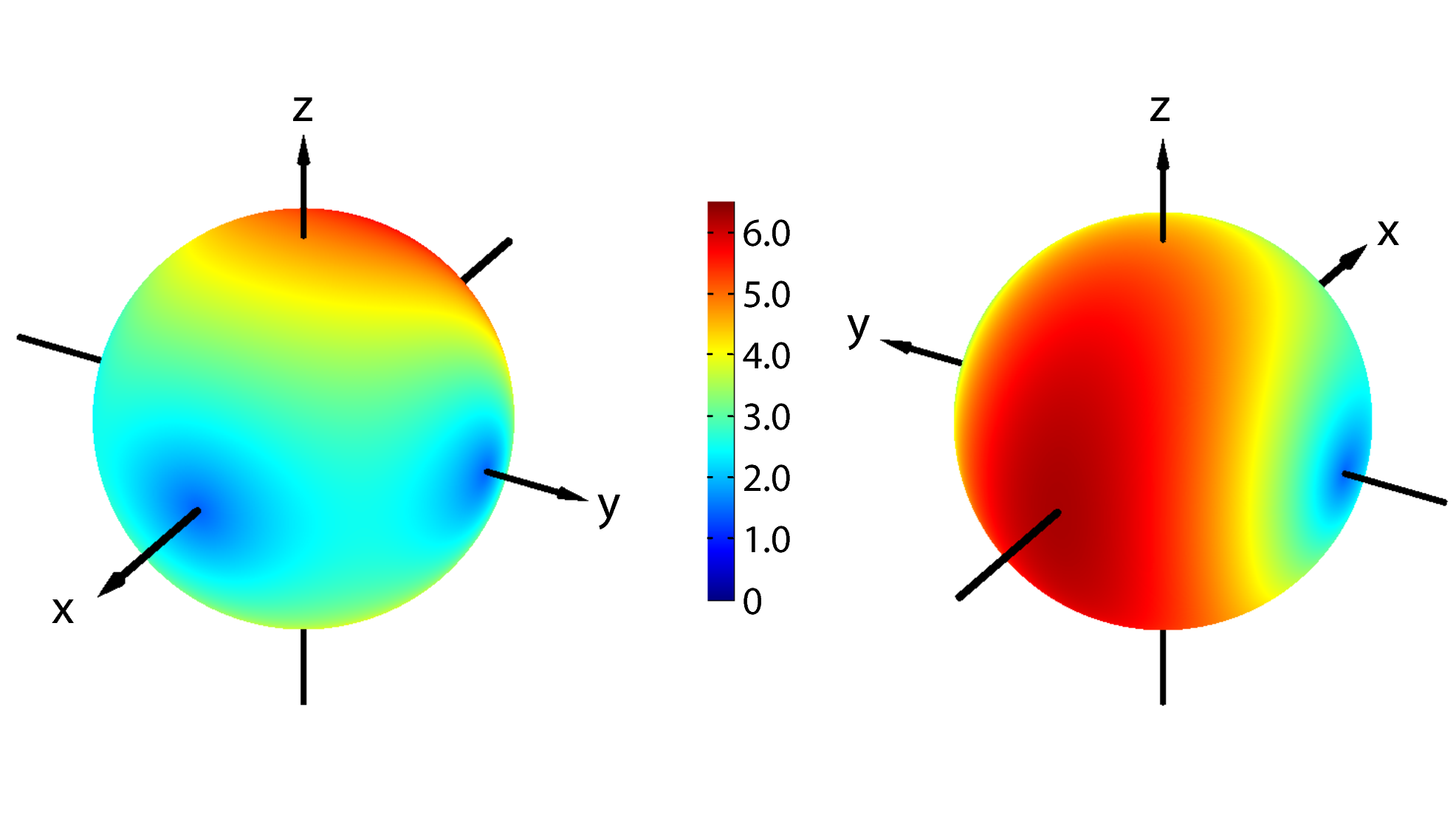}
\caption{(Color online) $\epsilon(A)\eta(B)+\epsilon(A)\sigma(B)+\sigma(A)\eta(B)$ for $\ket{\psi}=\ket{+z}$, $A=\sigma_x$ and $B=\sigma_y$. Every point on the Bloch sphere corresponds to a possible direction of $O_A$ and its color encodes the value of the sum of the three terms lying everywhere above $\frac{1}{2}|\bra{\psi}[A,B]\ket{\psi}|=1$. }
\label{fig:ozawa_bloch}
\end{center}
\end{figure}

For a complete investigation of Ozawa's error-disturbance uncertainty relation (\Eq{ozawa}) we additionally need the standard deviations of $A$ and $B$ for an arbitrary initial state $\ket{\psi}$ given by
\begin{equation}
\label{eq:standard_deviation_spins}
\sigma(A)=\sqrt{1- (\vec a\cdot\vec r)^2} \ , \quad
\sigma(B)=\sqrt{1- (\vec b\cdot\vec r)^2}
\end{equation}
For a graphical representation of the left hand side of the Ozawa relation \Eq{ozawa}, we have to fix the observables $A$ and $B$ and the state $\ket{\psi}$ and can then vary $O_A$ over the Bloch sphere and indicate the resulting values of the sum by different colors.
While the Heisenberg-term $\epsilon(A)\eta(B)$ is independent from the initial state, the sum in \Eq{ozawa} changes because of the standard deviations and always lies above the lower bound given by the commutator (see Fig.\,\ref{fig:ozawa_bloch}).

\section{Measurement Concept}
\label{sec:meas}

The expressions Eqs.\,\eq{error_Mm} and \eq{dist_Mm} are very convenient to calculate the explicit formulas for error and disturbance in spin measurements (Eqs.\,\eq{error_spin}, \eq{dist_spin}), but they are not suitable for designing an experiment. The output operator $O_A$ and the observable $A$ can not be measured simultaneously and therefore their difference is no detectable quantity. The same is true for the the change of the observable $B$ during the measurement. That's why error and disturbance have been claimed to be experimentally inaccessible \cite{Werner2004, Koshino2005}. But, this obstacle can be overcome by using the expressions Eqs.\,\eq{error_5terms} and \eq{dist_5terms} which read in our case of a projective spin-$\frac{1}{2}$ measurement
\begin{eqnarray}
\nonumber
\epsilon(A)^2&=&
2
+\bra{\psi}O_A\ket{\psi}
+\bra{A\psi} O_A \ket{A\psi}
\quad
\\
&&
-\bra{(A+\1)\psi} O_A \ket{(A+\1)\psi}
\label{eq:error_5terms_spin}
\end{eqnarray}
and
\begin{eqnarray}
\nonumber
\eta(B)^2&=&
2
+\bra{\psi}O_{B}\ket{\psi}
+\bra{B\psi} O_{B} \ket{B\psi}
\qquad\\
&&
-\bra{(B+\1)\psi} O_{B}\ket{(B+\1)\psi}
\label{eq:dist_5terms_spin}
\end{eqnarray}
since $A^2=O_A^{(2)}=B^2=O_{B}^{(2)}=\1$. Now, error and disturbance are solely determined by expectation values of experimentally accessible operators for various input states: 
\newline
The expression for the error only contains the output operator $O_A$, that is the observable that's actually measured by the apparatus. The desired observable $A$ occurs solely in two of the necessary three input states $\ket{\psi},\ket{A\psi}$,and $\ket{(A+\1)\psi}$. If we succeed in preparing all these known input states the error $\epsilon(A)$ of the $A$-measurement can be determined from the expectation values of $O_A$.

For determining the disturbance we need the input states $\ket{\psi},\ket{B\psi}$, and $\ket{(B+\1)\psi}$ and the expectation values of $O_B$ which is just a shorthand notation for $O_B^{(1)}$. From Eq.(\ref{eq:OB}) it is given by $\sum_{m}M_{m}^{\dagger} B M_{m}$ where the $M_{m}$ are the projection operators of the output operator $O_A$. Thus, if we measure $B$ immediately after the measurement of $O_A$ we actually perform the measurement of $O_B$.

\begin{figure}
\begin{center}
\includegraphics[width=8.5cm,keepaspectratio=true]{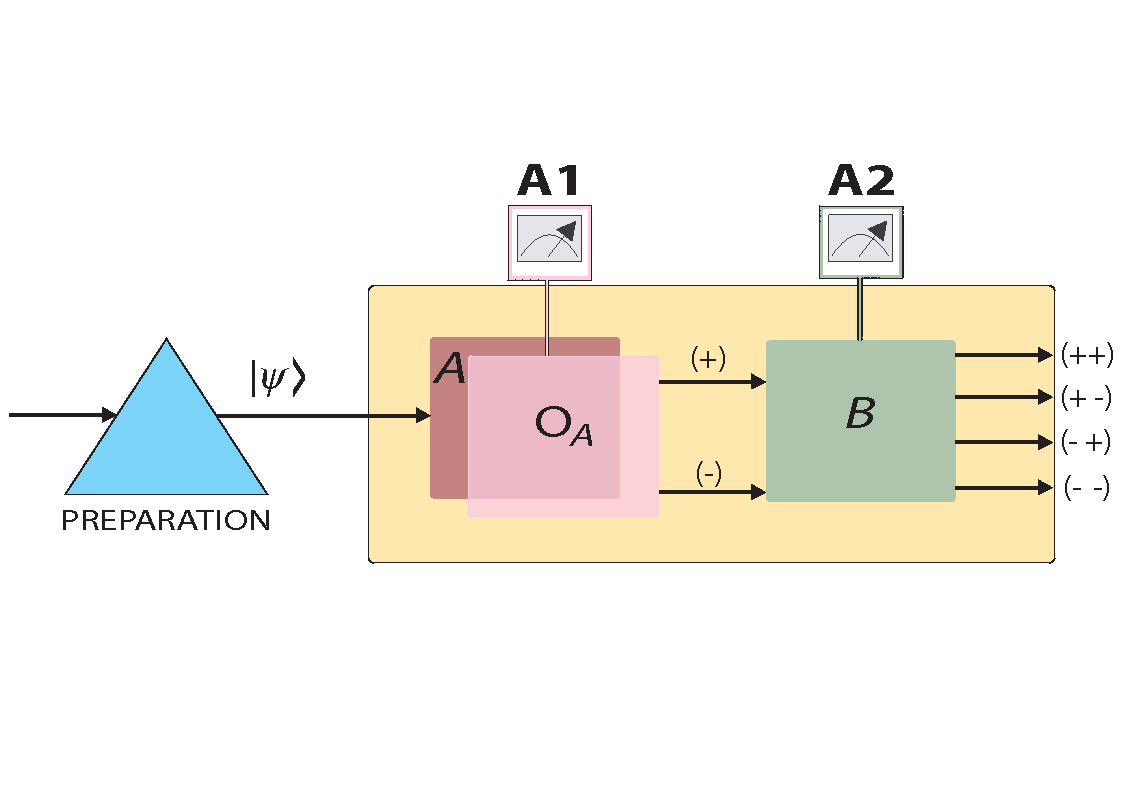}
\caption{(Color online) Experimental concept for the verification of the error-disturbance relation:
In the first measurement, the apparatus $\bAA$ is detuned in a way that it projectively measures the observable $O_A$ instead of $A$ thus causing an error $\epsilon$ in the $A$-measurement. The subsequent  measurement of observable $B$ in the eigenstate of $O_A$, performed by apparatus  $\bAAA$, virtually modifies $B$ to be $O_B$ from whose expectation values the disturbance $\eta$ on $B$ can be determined.}
\label{fig:measurement_scheme}
\end{center}
\end{figure}

After all, our experiment needs three connected components, a preparation stage for generating the input states $\ket{\psi}$, $\ket{A\psi}$, $\ket{(A+\1)\psi}$, $\ket{B\psi}$, and $\ket{(B+\1)\psi}$, a measurement apparatus $\bAA$ carrying out the $O_A$ measurement and a second apparatus $\bAAA$ performing the $B$ measurement (see Fig.\,\ref{fig:measurement_scheme}).

As indicated in the measurement scheme (Fig.\,\ref{fig:measurement_scheme}) the two projective measurement of $O_A$ and $B$ result in four possible outcomes since every spin operator can be decomposed into two projectors belonging to the eigenvalues $\pm 1$. Explicitly, these projectors are given by $\frac{1}{2}(\1\pm\vec o_a \cdot\vec \sigma)$ for $O_A$ and by $\frac{1}{2}(\1\pm\vec b \cdot\vec \sigma)$ for $B$. Equivalently, we can write the projector as ket-bra of the eigenvectors which we denote by $\ket{+\vec o_a}$ and $\ket{-\vec o_a}$ for the operator $O_A$ and by $\ket{+\vec b}$ and $\ket{-\vec b}$ for the operator $B$ respectively. We will use this kind of notation throughout the paper for all spin states. The vector inside the ket indicates the reference axis and the sign denotes the positive or negative spin component. For directions along the Cartesian axes we omit the arrow (for example $\ket{-x}$ stands for the negative spin component along the $x$-axis).

From the spectral theorem, the operators $O_A$ and $B$ can be decomposed into their projection operators via
\begin{eqnarray}
\label{eq:O_A_spectral}
O_A &=& \vec o_a\cdot\vec\sigma =
\ket{+\vec o_a}\bra{+\vec o_a}-\ket{-\vec o_a}\bra{-\vec o_a}\\
\label{eq:B_spectral}
B &=&\vec b\cdot\vec\sigma =
\ket{+\vec b}\bra{+\vec b}-\ket{-\vec b}\bra{-\vec b}
\end{eqnarray}
The indices $(++)$, $(+-)$, $(-+)$, and $(--)$ of the output ports in Fig.\,\ref{fig:measurement_scheme} indicate which projections have been carried out. From the intensities at the four possible output ports, denoted by $I_{++},I_{+-},I_{-+}$, and $I_{--}$, the expectation value of $O_A$ in a state $\ket{\psi}$ is obtained from
\begin{equation}
\label{eq:O_A_intensities}
\bra{\psi}O_A \ket{\psi} = \frac{(I_{++}+I_{+-})- (I_{-+}+I_{--})}{I_{++}+I_{+-}+I_{-+}+I_{--}}
\end{equation}
As already mentioned, the prior measurement of $O_A$ modifies the measurement operator of $\bAAA$ from $B$ to $O_{B}$. The expectation values of $O_B$ necessary for the determination of the disturbance are obtained from
\begin{equation}
\label{eq:X_B_intensities}
\bra{\psi}O_B \ket{\psi} = \frac{(I_{++}+I_{-+})- (I_{+-}+I_{--})}{I_{++}+I_{+-}+I_{-+}+I_{--}}
\end{equation}
%For a more explicit derivation how these expectation values arise from the experimental scheme we refer to appendix \ref{app:meas_concept}.
All expectation values necessary for the determination of error $\epsilon$ and disturbance $\eta$ can thus be derived from the output intensities when the input states $\ket{\psi}$, $\ket{A\psi}$, $\ket{(A+\1)\psi}$, $\ket{B\psi}$, and $\ket{(B+\1)\psi}$ are applied to the joint measurement apparatuses $\bAA$ and $\bAAA$.

Ozawa's relation (\Eq{ozawa}) additionally contains the standard deviations of $A$ and $B$ in the state $\ket{\psi}$. For our measurement observables, the standard deviations reduce to
\begin{equation}
\sigma(A)^2=\bra{\psi}A^2\ket{\psi}-\bra{\psi}A\ket{\psi}^2
=1-\bra{\psi}A\ket{\psi}^2
\end{equation}
and the corresponding expression for $\sigma(B)$. The expectation value of $A$ can be obtained from our experimental setup by adjusting $O_A$ to be equal to $A$ and then use \Eq{O_A_intensities}. For the expectation value of $B$, we have to turn off the measurement apparatus $\bAA$ and apply $\ket{\psi}$ to $\bAAA$ which then projectively measures $B$ (and not $O_B$). If we then denote the intensities of $\bAAA$'s two output ports by $I_+$ and $I_-$ the expectation value of $B$ is according to \Eq{B_spectral} given by $(I_+-I_-)/(I_+ + I_-)$. Obviously, for the expectation value of $A$ we can also turn off $\bAAA$ and only use the two output ports of $\bAA$ in the same manner. For the actual experiment we favor this method since increasing the number of involved components usually increases the experimental error.

A final remark concerns the principle aim of our experiment. We do not seek to minimize the measurement error $\epsilon(A)$ as one may assume at first sight. By the way, measuring the spin of a spin-$\frac{1}{2}$ particle along any direction is easily achievable with a high degree of precision. We are interested in a controlled variation of $O_A$ and a systematic investigation of the resulting measurement error $\epsilon(A)$ and disturbance $\eta(B)$, which are given by Eqs.\,\eq{error_5terms_spin} and \eq{dist_5terms_spin}, in order to demonstrate the behavior of Heisenberg's relation (\Eq{heisenberg_general}) in comparison to Ozawa's relation (\Eq{ozawa}).

\section{Neutron spin measurement setup}

\begin{figure*}
\begin{center}
\includegraphics[width=11.5cm,keepaspectratio=true]{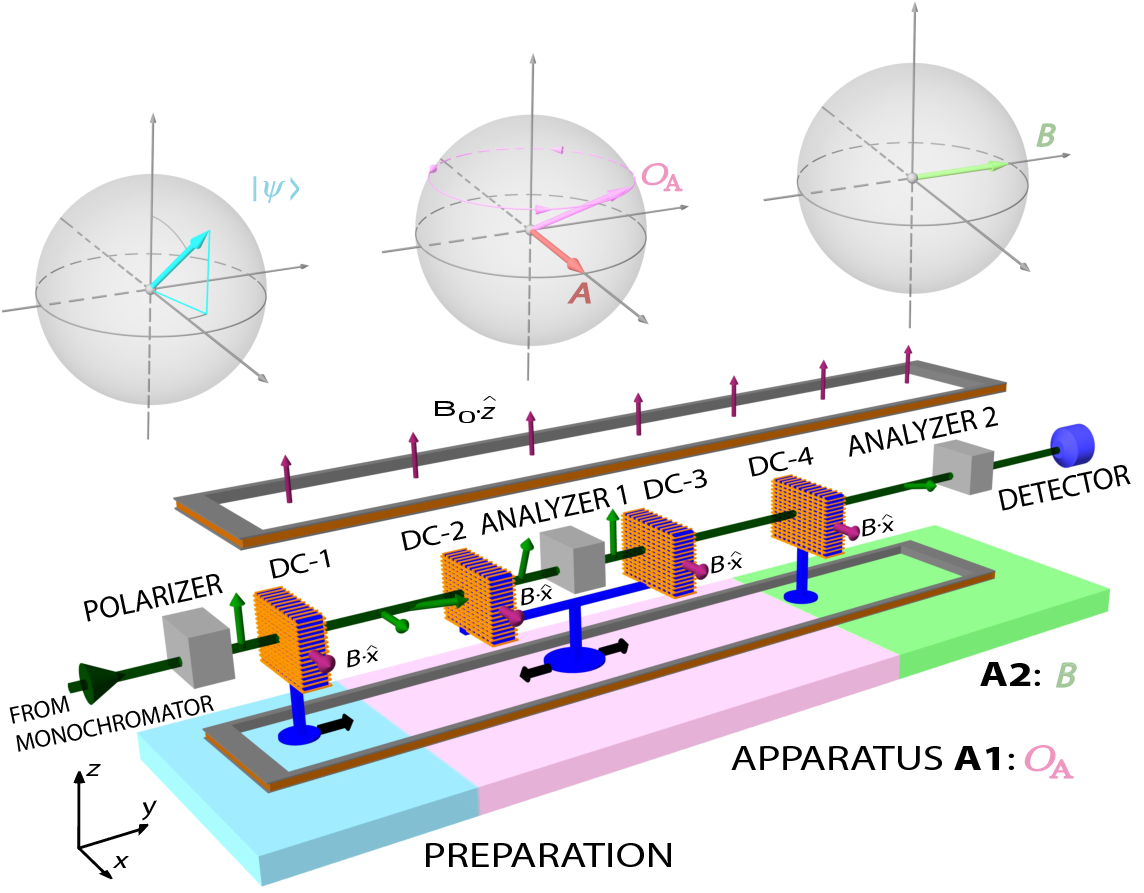}
\caption{(Color online) Illustration of the experimental apparatus. The set-up is designed for the demonstration of error and disturbance in neutron spin measurements.
The neutron optical set-up consists of three stages: preparation (blue region), apparatus $\bAA$ measuring $O_A$ (pink region) and apparatus $\bAAA$ measuring $B$ (green region). A monochromatic neutron beam is polarized in $+z$-direction by passing through a supermirror spin polarizer. By combining the action of four spin-flipper coils, the magnetic guide field and the analyzing supermirrors, the successive measurements of $O_A$ and $B$ are made for the required input states.
}
\label{fig:measurement_setup}
\end{center}
\end{figure*}

In the previous section, we have outlined the idea behind the experiment. We now want to turn to the actually performed measurements on neutron spins. The experiment was carried out at the tangential beam port of the research reactor facility TRIGA Mark II of the Vienna University of Technology.
The whole setup is depicted in Fig.\,\ref{fig:measurement_setup}.

A monochromatic, thermal neutron beam with a mean wavelength of 1,96 $\AA$ and a cross section of about 10 (vertical) $\times$ 5 (horizontal) mm$^2$
propagates in the $y$-direction. In the so-called preparation stage, it first crosses a bent Co-Ti supermirror resulting in an approximately 99$\%$ polarization in $+z$-direction. For the further manipulation of the spin state we use magnetic fields which interact with the neutron via the Zeeman Hamitonian $\mu \vec{\sigma}\vec B$. For thermal neutrons the effect of the magnetic fields on the spatial part of the wave function can be neglected \cite{Klepp08} and the action of a static magnetic field pointing in direction $\vec n_{B}$ on the spin state is described by a unitary transformation $U_R$
\begin{equation}
U_R=e^{i  \alpha \ \vec n_{B} \cdot \vec{\sigma}},
\quad\alpha=\mu |B| T/\hbar
\end{equation}
where $\mu$ denotes the magnetic moment of the neutron, $|B|$ the modulus of the magnetic field strength, and $T$ the time-of-flight through the field region. The magnetic field thus induces a rotation of the neutron polarization around the axis $\vec n_B$ with an angle $\alpha$, commonly referred to as Larmor precession. In order to obtain any arbitrary spin state we combine the action of a 10 Gauss guide field pointing in $z$-direction and so-called spin-flippers. The spin flipper produces a field in negative $z$-direction which compensates the guide field and a second field $B_x$ that points in $x$-direction. When the $z$-polarized neutrons propagate through the spin flipper their polarization obtains a polar angle depending on the strength of $B_x$. In the guide field, Larmor precession around the $z$-axis is induced. Since the strength of the guide field is fixed, the azimuthal angle of the polarization is varied by changing the time-of-flight through the guide field. Thus, in order to get a certain azimuthal angle at the end of the preparation stage, we have to position the first spin flipper DC-1 properly between the polarizer and the end point of the preparation stage.

Carrying out the projective measurement of $O_A$ consists of two steps. At first, we have to project the initially prepared state onto the eigenstates of $O_A$. To complete the measurement we then have to prepare the neutron spin in the eigenstates of $O_A$. For the projection, spin-flipper DC-2 has to be positioned in the guide field such that the spin-component along the axis to be measured is rotated into the $+z$-direction. For the eigenstate belonging to eigenvalue $+1$, we need the component along $\vec o_a$, for the eigenstate belonging to $-1$, we rotate the $-\vec o_a$ component in $+z$-direction. The supermirror (analyzer) then selects only the $\ket{+z}$-part of the wave function. The projective measurement is completed by the preparation of the measured spin-component with spin-flipper DC-3. In analogous manner to the preparation of the initial state, this is accomplished by properly positioning spin flipper DC-3 in the guide field, so that the desired state is generated at the exit of apparatus $\bAA$.

Note that the magnetic guide field strength and the dimensions of our experiment are chosen such that any desired direction for the output operator $O_A$ and any initial state can be realized.

On the state after the $O_A$ measurement, the $B$-measurement is performed using the last spin-flipper (DC-4) and the analyzer. A subsequent preparation of the eigenstates of $B$ is not necessary since the counting detector is insensitive to the spin state.

\begin{figure}[!htb]
\begin{center}
\includegraphics[width=\columnwidth,keepaspectratio=true]{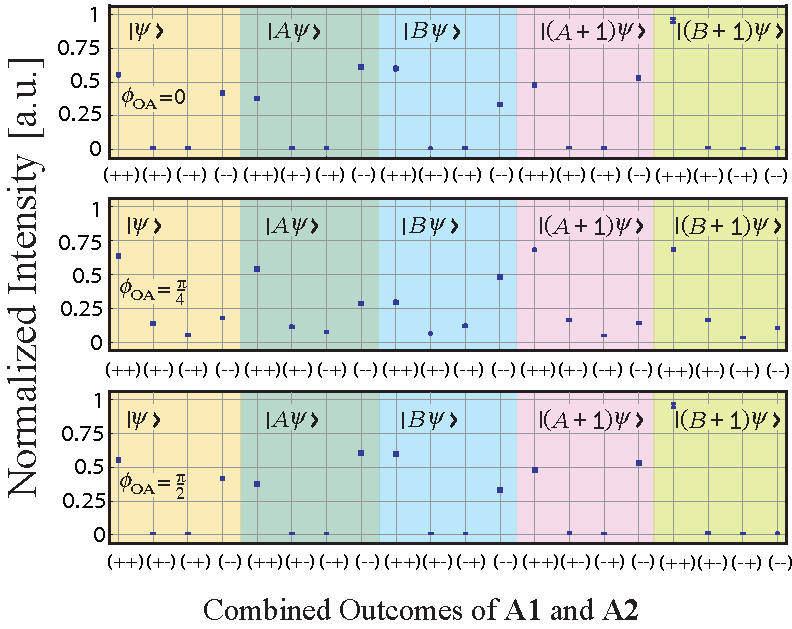}
\caption{(Color online) Normalized intensity of the successive measurements carried out by apparatus $\bAA$ and $\bAAA$. The combined projective measurements of $O_A$ and $B$ have four outcomes, denoted as ($+\,+$), ($+\,-$), ($-\,+$) and ($-\,-$) and have to be recorded for each initial state, i.e. $\ket{\psi}$, $\ket{A\psi}$, $\ket{B\psi}$, $\ket{(A+\1)\psi}$ and $\ket{(B+\1)\psi}$. Here $A=\sigma_x$, $B=\sigma_y$ and $\ket{\psi}=\ket{\theta_\psi=\pi/4,\phi_\psi=\pi/12}$. $O_A$ is varied within the $xy$-plane with azimuthal angle given by $\phi_{OA}=0,\pi/4$ and $\pi/2$. Error bars represent $\pm$ one standard deviation of the normalized intensities. Some error bars are at the size of the markers.}
\label{fig:intensitiesAB}
\end{center}
\end{figure}

In contrast to an ideal measurement, in our setup the four possible outcomes of the two projective spin measurements along $\vec o_a$ and $\vec b$ are not measured at the same time, but one after the other. At first, we adjust apparatus to measure the projection onto the positive eigenstates $\ket{+o_a}$ and $\ket{+b}$ of $O_A$ and $B$. The resulting intensity derived from the counts/600 sec is consequently denoted by $I_{++}$. Afterwards, we change apparatus $\bAAA$ to constitute the projection operator $\ket{-b}\bra{-b}$ yielding $I_{+-}$. Then we change apparatus $\bAA$ to measure $\ket{-o_a}\bra{-o_a}$ and repeat the procedure for both eigenstates of $B$ to get $I_{-+}$ and $I_{--}$. From the four output intensities, we obtain the expectation values of $O_A$ and $O_B$ as given by Eqs.\,\eq{O_A_intensities} and \eq{X_B_intensities}. The measured expectation values are normalized taking the limiting efficiency of the entire apparatus ($\sim96\%$) into account, so that expectation values are bounded between $\pm 1$. In order to get results for error $\epsilon$ and disturbance $\eta$ the expectation values have to be recorded for the different states $\ket{\psi}$, $A\ket{\psi}$, $(A+\1)\ket{\psi}$, $B\ket{\psi}$, and $(B+\1)\ket{\psi}$. In Fig.\ref{fig:intensitiesAB}, we show an explicit example of related intensity sets.

For special configurations of $A$, $B$, and $\ket{\psi}$ some of these necessary states are equivalent since overall phase factors are irrelevant. For example if $A=\sigma_x$, $B=\sigma_y$ and $\ket{\psi} = \ket{+z}$ we have $\ket{A\psi}=\ket{-z}$ and $\ket{B\psi}=i\ket{-z}$.

For the measurement of the standard deviation $\sigma(A)$, we chose $O_A$ to be $A$ and then project the initial state $\ket{\psi}$ onto the eigenstates of $A$. A subsequent preparation of the eigenstates and the $B$-measurement are not necessary, that means the spin flippers DC-3 and DC-4 can be turned off. For the standard deviation $\sigma(B)$, we virtually remove apparatus $\bAA$ by turning off the spin flipper coils DC-1 and DC-2 and then use DC-3 (in combination with the guide field) to prepare the state $\ket{\psi}$ that is then applied to apparatus $\bAAA$ that projectively measures the expectation value of $B$.

With known error $\epsilon(A)$, disturbance $\eta(B)$, and standard deviations $\sigma(A)$ and $\sigma(B)$ we can examine the behaviour of Heisenberg's (\Eq{heisenberg_general}) and Ozawa's (\Eq{ozawa}) relation for varying $O_A$.

\section{Experimental Results}

\subsection{Standard configuration}

In the first experiment, we investigate the uncertainty relations in the so-called standard configuration where the Bloch vectors of the observables and the initial state ($\vec a$, $\vec b$, and $\vec r$) are orthogonal to each other
\begin{equation}
A=\sigma_x, \quad B=\sigma_y,\quad \ket{\psi}=\ket{+z}
\end{equation}
Note that since all results only depend on the relative orientation of the involved quantities we can always fix one direction. Therefore, we will chose $A$ to be $\sigma_x$ in all experiments.
For the standard configuration, the standard deviations and the expectation value of commutator 
between $A$ and $B$ become maximal (see Eqs.\,\eq{standard_deviation_spins} and \eq{comutator_spins})
\begin{equation}
\label{eq:sigma_limit_std_config}
\sigma(A)=1, \quad \sigma(B)=1,\quad \frac{1}{2}\bra{\psi}[A,B]\ket{\psi}=1
\end{equation}
In sec.\ref{sec:theory_spin} ("Theory for spin measurements"), we have depicted $\epsilon(A)$ (Fig.\,\ref{fig:error_bloch}), $\eta(B)$ (Fig.\,\ref{fig:dist_bloch}), their product $\epsilon(A)\eta(B)$ (Fig.\,\ref{fig:errordist_bloch}), and the sum
$\epsilon(A)\eta(B)+\epsilon(A)\sigma(B)+\sigma(A)\eta(B)$ (Fig.\,\ref{fig:ozawa_bloch}) for $A=\sigma_x$ and $B=\sigma_y$ for all possible directions of $O_A$ on the Bloch sphere. In our first experimental run, we vary $O_A$ along the equator. $O_A$ is thus parameterized by its azimuthal angle $\phi_{OA}$ yielding the direction $\vec o_a = (\cos \phi_{OA},\sin \phi_{OA},0)$. The theory curves for $\epsilon(A)$ and $\eta(B)$ are then given by
\begin{equation}
\label{eq:error_dist_std_config}
\epsilon(A) = 2 \sin \frac{\phi_{OA}}{2},\quad \eta(B) = \sqrt 2 |\cos \phi_{OA}|
\end{equation}
which follows from Eqs.\,\eq{error_spin} and \eq{dist_spin}.

In the standard configuration, we have to prepare $\ket{+z}$, $\ket{-z}$, $\ket{+x}$, and $\ket{+y}$ as input states in order to obtain the necessary expectation values that determine error, disturbance and standard deviations.
The errors of the measured values for $\epsilon(A)$ and $\eta(B)$ are calculated
using error propagation from the standard deviation of the count
rates and considering inaccuracies of the Larmor precession angles
($\sim 1,5^{\circ}$). The latter result mainly from inhomogeneity of the guide
field along the beam.

\begin{figure}[!htb]
\begin{center}
\includegraphics[width=\columnwidth,keepaspectratio=true]{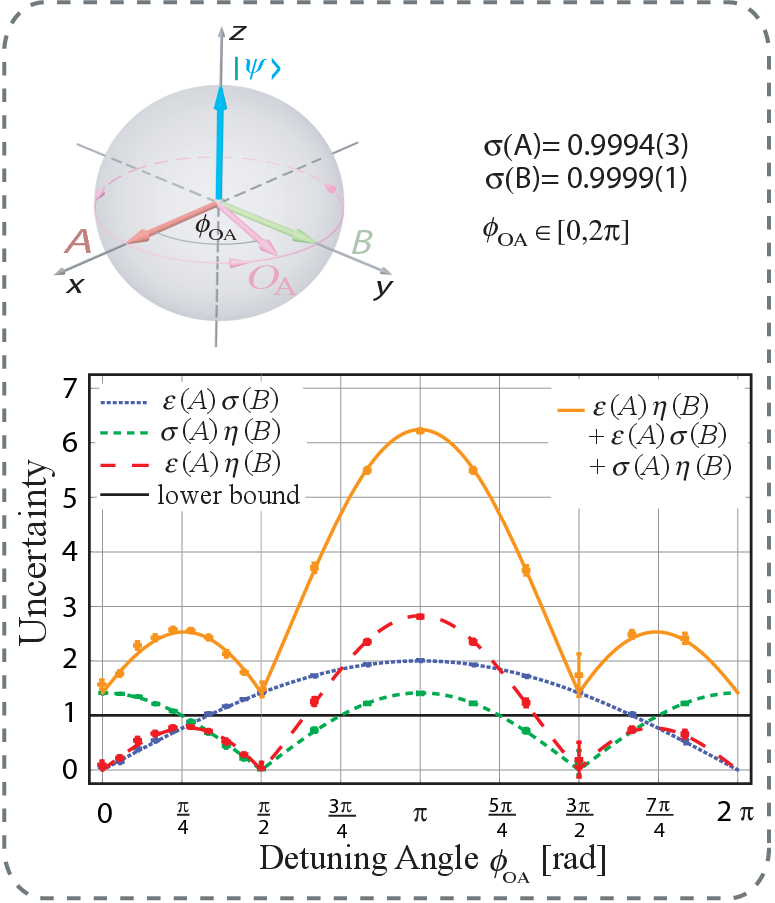}
\caption{(Color online) Experimentally determined values of error $\epsilon(A)$, disturbance $\eta(B)$, $\sigma(A)$, and $\sigma(B)$ plotted as $\epsilon(A)\sigma(B)$, $\sigma(A)\eta(B)$, $\epsilon(A)\eta(B)$ (this term corresponds to the left-hand-side 
of the Heisenberg relation (\Eq{heisenberg_general}) and 
as the sum $\epsilon(A)\sigma(B)+\sigma(A)\eta(B)+\epsilon(A)\eta(B)$  
(corresponds to the left-hand-side of the Ozawa's relation (\Eq{ozawa}) against the azimuthal angle of $O_A$. The observables $A$ and $B$ and the initial state $\ket{\psi}$ are depicted on the Bloch sphere, as well as the path along which $O_A$ is varied. The respective theory curves are given by Eqs.\,\eq{sigma_limit_std_config},\eq{error_dist_std_config}.}
\label{fig:standard_config}
\end{center}
\end{figure}

In Fig.\,\ref{fig:standard_config}, we show the experimental outcomes for the three terms occurring in Ozawa's relation plotted against the azimuthal angle of $O_A$ which we call detuning angle since it also indicates the amount of deviation between $A$ and $O_A$.
The standard deviations are independent of $O_A$ and their measured values are practically equal to $\sigma(A)=\sigma(B)=1$, that's why we can also investigate the behaviour of $\epsilon(A)$ from the $\epsilon(A)\sigma(B)$-curve and likewise the behaviour of $\eta(B)$ from the $\sigma(A)\eta(B)$-curve.

Initially, for $\phi_{OA}=0$, that is $O_A=A$, the measurement error $\epsilon(A)$ vanishes. When we move along the equator it increases and reaches its maximal value for $O_A=-A$ ( $\phi_{OA}=\pi$). Then, $O_A$ approaches $A$ again and $\epsilon(A)$ decreases.

The disturbance is maximal for $\phi_{OA}=0$ and vanishes when $O_A=B$ ($\phi_{OA}=\pi/2$). It has a second maximum for $O_A=-A$ because a $(-A)$-measurement disturbs $B$ as much as an $A$-measurement. $A$ and $-A$ have the same set of eigenstates and therefore the same set of projectors which leads already for the general formula \Eq{dist_Mm} to an equal expression for the disturbance. We can also verify that the disturbance vanishes again for $O_A=-B$ ($\phi_{OA}=3\pi/2$) which follows from similar arguments for $B$ and $-B$.

The famous trade-off relation stating that when one observable
is measured more precisely, the other is more disturbed has to be treated with care. This reciprocity between error and disturbance only holds for $-\pi/2\leq\phi_{OA}\leq\pi/2$. When we move away from $\phi_{OA}=\pi/2$ where $\eta(B)=0$ to increasing $\phi_{OA}$ we again disturb the $B$-measurement, but at the same time the error of the $A$-measurement increases as well since we tend towards $-A$. Then, for $\pi\leq\phi_{OA}\leq 3\pi/2$ both decrease. 

The product of error and disturbance $\epsilon(A)\eta(B)$ lies below the limit given by the commutator for the majority of $\phi_{OA}$-values revealing the violation of the generalized Heisenberg relation (\Eq{heisenberg_general}). Strongest violation occurs around the regions of error-free ($\phi_{OA}=0$) or disturbance-free ($\phi_{OA}=\pi/2, 3\pi/2$) measurements.

Contrary to that, the sum $\epsilon(A)\eta(B)+\epsilon(A)\sigma(B)+\sigma(A)\eta(B)$ is always above the expectation value of the commutator showing the validity of the Ozawa's relation (\Eq{ozawa}).

\subsection{Arbitrary direction of $O_A$}

\begin{figure*}
\begin{center}
\includegraphics[width=17cm,keepaspectratio=true]{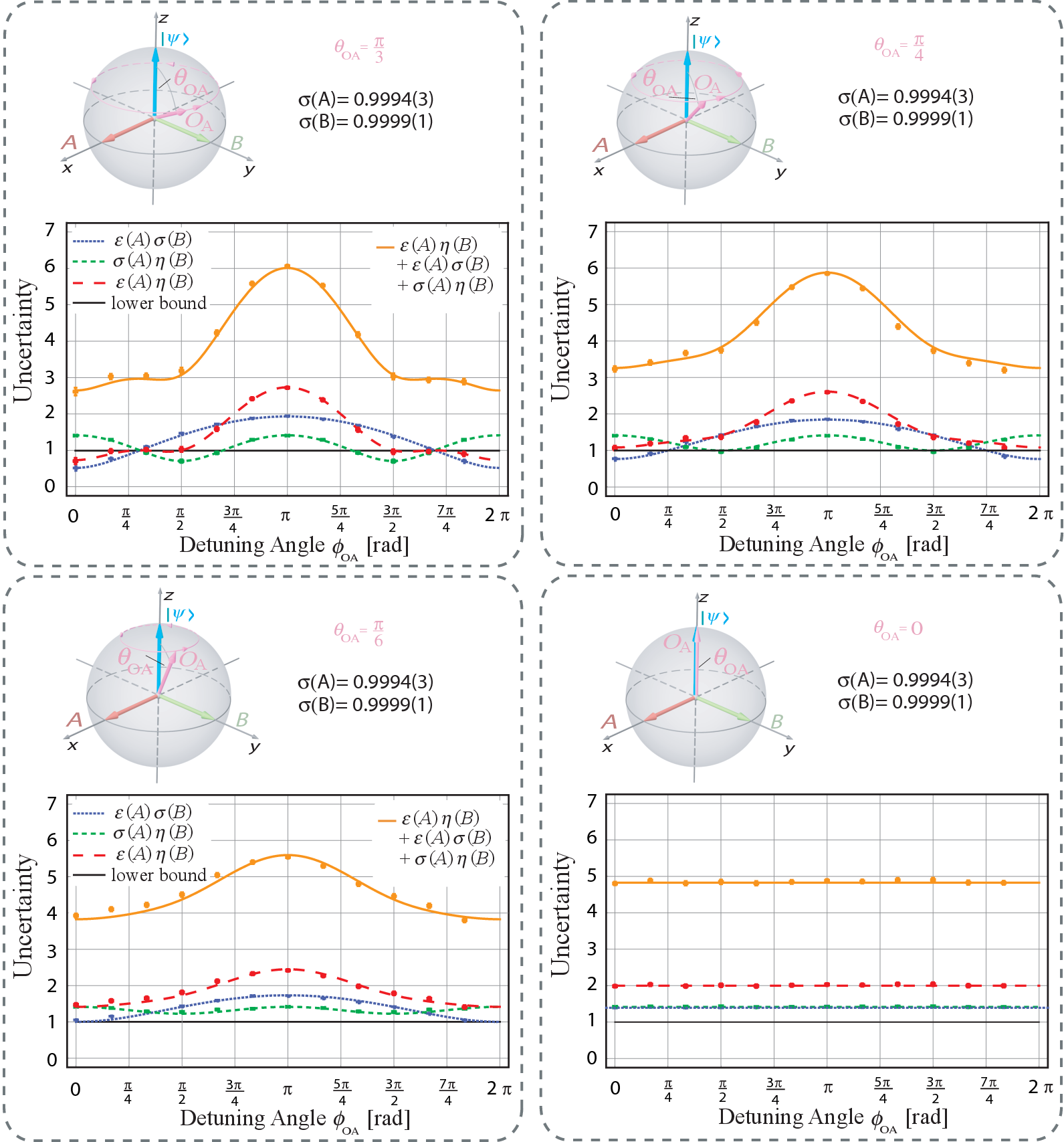}
\caption{(Color online) Standard configuration results for $O_A$ outside
$xy$-plane. The observables $A$ and $B$, the initial state $\ket{\psi}$ and the chosen path for the output operator $O_A$ are depicted on the Bloch sphere. The theoretically predicted values for $\sigma(A)$, $\sigma(B)$ and the lower limit are given by \Eq{sigma_limit_std_config} and the theory curves for $\epsilon(A)$ and $\eta(B)$ by \Eq{error_thetaOA} and \Eq{dist_thetaOA} respectively.} 
\label{fig:standard_config_thetaOA}
\end{center}
\end{figure*}

Though moving $O_A$ along the equator of the Bloch sphere already reveals a lot of remarkable features about error, disturbance and the related inequalities a systematic investigation requires arbitrary variations of $O_A$. The direction of $O_A$ is in general given by an azimuthal angle $\phi_{OA}$ and a polar angle $\theta_{OA}$, so that $\vec o_a =(\cos \phi_{OA}\sin \theta_{OA},\sin \phi_{OA}\sin \theta_{OA},\sin \theta_{OA})^T$ which yields for error and disturbance if still $A=\sigma_x$ and $B=\sigma_y$
\begin{eqnarray}
\label{eq:error_thetaOA}
\epsilon(A) &=& \sqrt{2-2\cos\phi_{OA}\sin\theta_{OA}}\\
\label{eq:dist_thetaOA}
\eta(B) &=& \sqrt{2-2\sin^2\phi_{OA}\sin^2\theta_{OA}}
\end{eqnarray}
For the following experiments, we will fix the polar angle $\theta_{OA}$ and then vary the azimuthal angle $\phi_{OA}$ thereby realizing an evolution of $O_A$ along circles of latitude on the Bloch sphere (see Fig.\,\ref{fig:standard_config_thetaOA}).

Neither error nor disturbance can then vanish completely since $O_A$ never coincides with $A$, $B$ or $-B$ along these paths. The $\epsilon$- and $\eta$-curves are as a whole shrunken from below. Note that we can deduce the behaviour of $\epsilon(A)$ and $\eta(B)$ from Fig.\,\ref{fig:standard_config_thetaOA} since $\sigma(A)$ and $\sigma(B)$ are still 1. The shrinking effect increases the more $\theta_{OA}$ differs from $\pi/2$ ending up in straight lines for the degenerate case $\theta_{OA}=0$.

The trade-off effect between error and disturbance is weakened, but still valid for $-\pi/2\leq\phi_{OA}\leq\pi/2$ if we move along the circles of latitude. If we move along meridians no trade-off exists at all since error and disturbance both increase. Thus, at least for spin measurements, the trade-off is not only restricted to certain regions but also to certain directions in the parameter space.

Since both error and disturbance increase their product increases as well and lies above the limit in larger and larger intervals. For $\theta_{OA}\leq \arcsin(0.75)\approx 48.6^{\circ}$ the Heisenberg relation is always fulfilled (see for example top right panel of Fig.\ref{fig:standard_config_thetaOA}).

Ozawa's inequality is again always fulfilled, the sum $\epsilon(A)\eta(B)+\epsilon(A)\sigma(B)+\sigma(A)\eta(B)$ gets shifted far above the limit when $O_A$ approaches the poles.

\subsection{Varying the azimuthal angle of $B$}

\begin{figure*}
\begin{center}
\includegraphics[width=17cm,keepaspectratio=true]{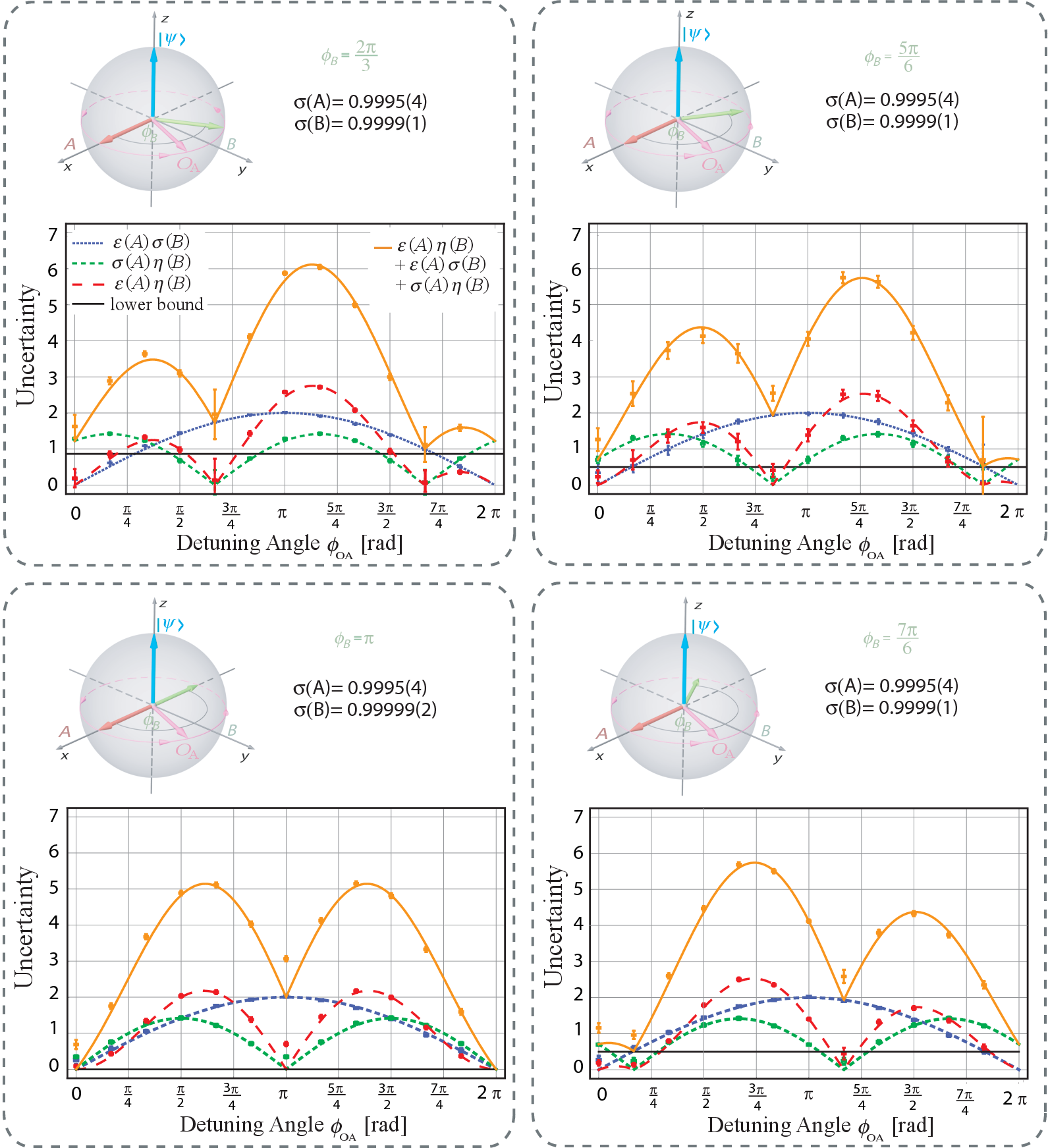}
\caption{(Color online) Variations of the azimuthal angle $\phi_B$ of the observable $B$.  The observables $A$ and $B$, the initial state $\ket{\psi}$ and the chosen path for the output operator $O_A$ are depicted on the Bloch sphere. The theoretically calculated values for $\sigma(A)$, $\sigma(B)$ and the lower limit are given by \Eq{sigma_limit_phiB} and the theory curves for $\epsilon(A)$ and $\eta(B)$ by \Eq{error_dist_phiB}.} \label{fig:phiB_variations}
\end{center}
\end{figure*}

In the next step we quit the standard configuration and change the relative orientation of $A$ and $B$ by moving $B$ along the equator (see Fig.\,\ref{fig:phiB_variations}).

\begin{equation}
A=\sigma_x, \quad B=\sigma_x \cos\phi_B +\sigma_y \sin\phi_B,\quad \ket{\psi}=\ket{+z}
\end{equation}
which leads to
\begin{equation}
\label{eq:sigma_limit_phiB}
\sigma(A)=1, \quad \sigma(B)=1,\quad
\frac{1}{2}\bra{\psi}[A,B]\ket{\psi}=|\sin \phi_B|
\end{equation}
If we also vary $O_A$ along the equator the error remains unaltered
compared to the standard configuration but we expect the disturbance
curve to be shifted an amount $\phi_B-\pi/2$
\begin{equation}
\label{eq:error_dist_phiB}
\epsilon(A) = 2 \sin \frac{\phi_{OA}}{2},\quad \eta(B) = \sqrt 2 |\sin( \phi_{OA}-\phi_B)|
\end{equation}
Note that in the standard configuration $\phi_B=\pi/2$ which turns the sine dependence into a cosine.

The error still vanishes for $O_A=A$ and has its maximum at $O_A=-A$, the disturbance vanishes for $O_A=B$ and $O_A=-B$ and has maxima in the middle between the two minima which do not correspond to the direction of $A$ or $-A$ now.

Due to the loss of symmetry, the product $\epsilon(A)\eta(B)$ changes considerably. Since the commutator between $A$ and $B$ additionally decreases, compared to the standard configuration, the regions where the Heisenberg error-disturbance relation is violated become smaller.

The expression $\epsilon(A)\eta(B) + \epsilon(A)\sigma(B) + \sigma(A)\eta(B)$ is always above the limit again demonstrating the validity of Ozawa's relation. However, in the standard configuration the sum never touched the limit whereas for shifted $B$ it approaches the lower bound. Points associated with disturbance-free measurement come closest to the limit and finally touch it for the degenerate cases $B=A$ and $B=-A$. 

In this context we want to mention an interesting aspect concerning the lower bound. For the pre-measurement standard deviations in the state $\ket{\psi}$, the lower bound derived by Robertson \cite{Robertson29} and given by \Eq{robertson} was soon refined by Schr\"odinger \cite{Robertson_Schroedinger} to be
\begin{equation}
\label{eq:robertson_schroedinger}
\sigma(A)^2 \cdot\sigma(B)^2 \ge
\langle \frac{1}{2}\{ A - \langle A \rangle, B-\langle B\rangle \}\rangle^2 +
\langle \frac{1}{2i}[ A , B]\rangle^2
\end{equation}
where $\langle..\rangle$ abbreviates $\bra{\psi}..\ket{\psi}$ and $\{X,Y\}=XY+YX$ stands for the anti-commutator. For our observables $A=\vec a \cdot \vec\sigma$, $B=\vec b \cdot\vec\sigma$ and the initial state $\ket{\psi}\bra{\psi} = \frac{1}{2}(\1 + \vec r \cdot\vec\sigma)$ the additional term explicitly reads
\begin{equation}
\label{eq:anti_commutator_spin}
\frac{1}{2}\langle\{ A - \langle A \rangle, B-\langle B\rangle \}\rangle
= \vec a\cdot\vec b - (\vec a\cdot \vec r) (\vec b\cdot \vec r) 
\end{equation}
It vanishes for the standard configuration, but for non-orthogonal $A$ and $B$ it contributes. The measured standard deviations in Fig.\,\ref{fig:phiB_variations} of course obey the improved relation \Eq{robertson_schroedinger}. But for the error-disturbance relations, the anti-commutator term can not be added to the lower bound. This shows explicitly that the three terms of Ozawa's relation are not bounded from below by the product of the pre-measurement standard deviations as one could naively conclude from the results of the standard configuration.

When moving $O_A$ along the equator we find out that the trade-off relation between error and disturbance is valid in unconnected intervals for $\phi_{OA}$ whose positions depend on the orientation of $B$.

\subsection{Varying the polar angle of $B$}

Now, we also introduce a polar angle for $B$ while leaving its azimuthal angle to be $\frac{\pi}{2}$. The observable and the initial state are then given by
\begin{equation}
A=\sigma_x, \quad B=\sigma_y \sin \theta_B +\sigma_z \cos\theta_B,\quad \ket{\psi}=\ket{+z}
\end{equation}
which leads to
\begin{equation}
\label{eq:sigma_limit_thetaB}
\sigma(A)=1, \quad \sigma(B)=\sin\theta_B,
\quad \frac{1}{2}\bra{\psi}[A,B]\ket{\psi}=\sin \theta_B
\end{equation}
We vary $O_A$ along the equator and along a circle of latitude determined by $\theta_{OA}=\frac{\pi}{3}$. The error $\epsilon(A)$ remains the same as in \Eq{error_thetaOA} whereas the expression for the disturbance written in spherical coordinates becomes rather lengthly and we refer to the general expression \Eq{dist_spin}.

\begin{figure}[!hbt]
\begin{center}
\includegraphics[width=8.5cm,keepaspectratio=true]{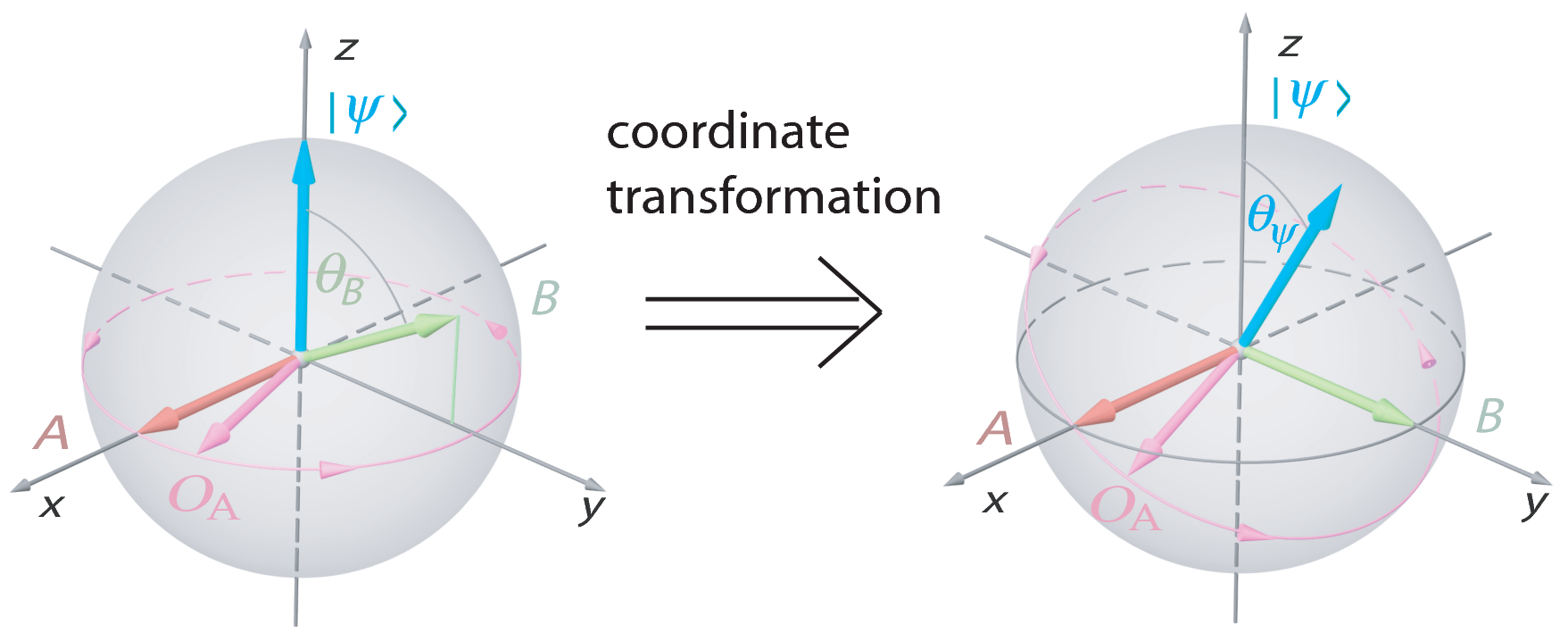}
\caption{(Color online) Since only the relative orientations of $A$, $B$, $\ket{\psi}$, and $O_A$ determine the results, the experiments involving the polar angle $\theta_B$ can be equivalently seen as configurations with a changed initial state and an evolution of $O_A$ along an inclined plane.}
\label{fig:theta_B_rotation}
\end{center}
\end{figure}

Note that we can obviously perform a coordinate transformation so that $A$ and $B$ lie in the $xy$-plane again (see Fig.\,\ref{fig:theta_B_rotation}). Then, the state $\ket{\psi}$ gets a polar angle and the path of $O_A$ inclines likewise and consequently its parametrization becomes more elaborate. Nevertheless, one can also see the above experiments as a realization of this particular evolution of $O_A$ for $A=\sigma_x$, $B=\sigma_y$ and a declined initial state.

\begin{figure*}
\begin{center}
\includegraphics[width=\textwidth,keepaspectratio=true]{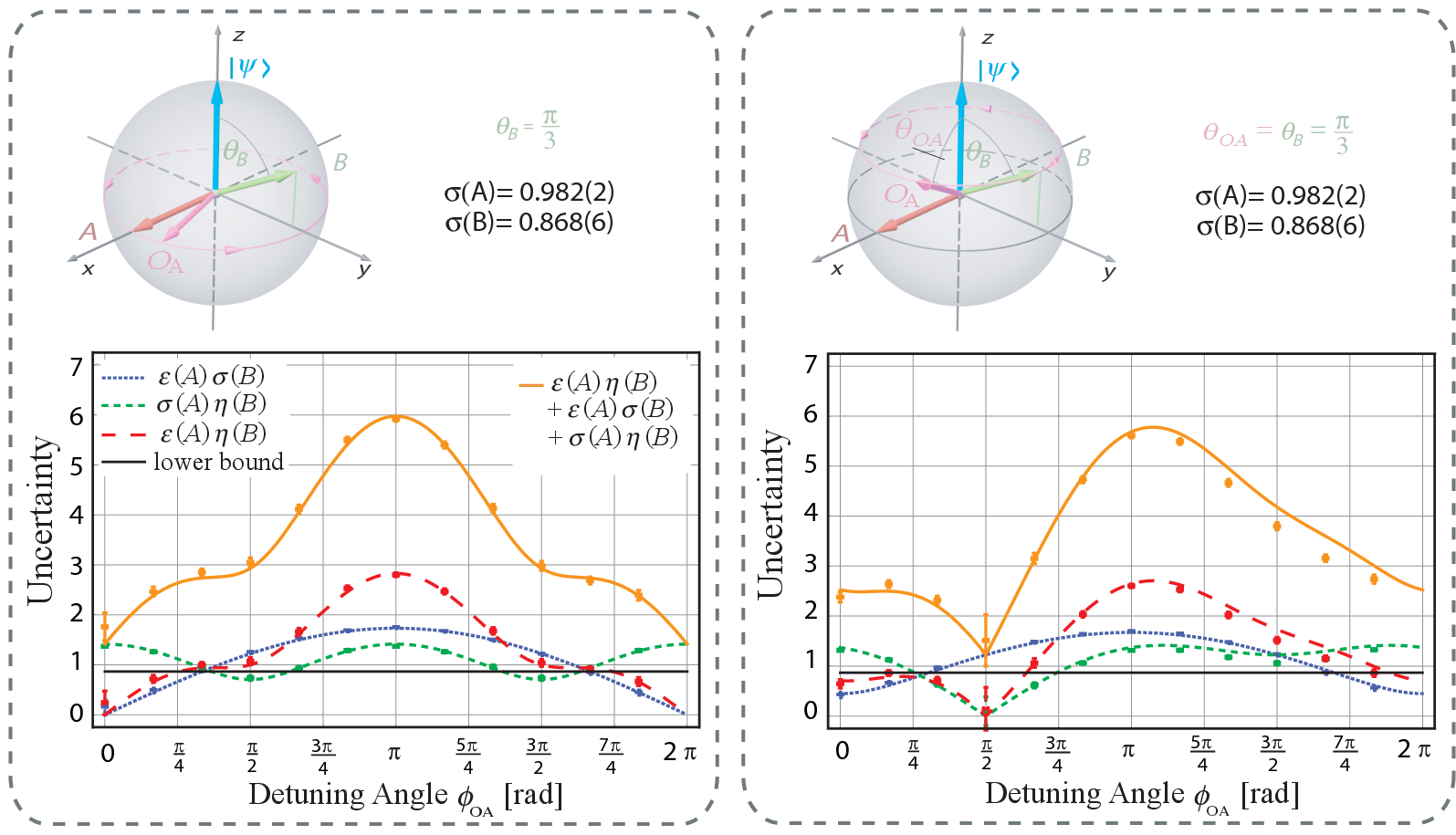}
\caption{(Color online) Variations of the polar angle $\theta_B$ of the observable $B$. The observables $A$ and $B$, the initial state $\ket{\psi}$ and the chosen path for the output operator $O_A$ are depicted on the Bloch sphere. The theoretically predicted values for $\sigma(A)$, $\sigma(B)$ and the lower limit are given by \Eq{sigma_limit_thetaB} and the theory curves for $\epsilon(A)$ and $\eta(B)$ by \Eq{error_thetaOA} and \Eq{dist_spin}, respectively.}
\label{fig:thetaB}
\end{center}
\end{figure*}

For the $O_A$-variation along the equator, the error curve does not change compared to the standard configuration, the term $\epsilon(A)\sigma(B)$ in Fig.\,\ref{fig:thetaB} becomes smaller due to smaller $\sigma(B)$. The disturbance never vanishes totally since $O_A$ never coincides with $B$ or $-B$. The curve is still symmetric because the enclosed angles between $O_A$ and $B$ for $0\leq\phi_{OA}<\pi$ are the same as the angles between $O_A$ and $-B$ for $\pi\leq\phi_{OA}<2\pi$. 

When $O_A$ varies along the circle of latitude given by $\theta_{OA}=\theta_B$ the situation changes. The error can not become zero since $O_A\neq A$ holds everywhere. The disturbance vanishes for $\phi_{OA}=\frac{\pi}{2}$ where $O_A=B$ but the symmetry is lost. $O_A$ does not approach $-B$ in the same manner as $B$ and so the disturbance is hardly reduced for $\pi\leq\phi_{OA}<2\pi$. 

\subsection{Varying the initial state}

Until now, we have varied $O_A$ along different paths and changed the direction of $B$. The next thing to investigate are alterations of the initial state $\ket{\psi}$ for an already examined choice of $A$ and $B$ which should, according to the theoretical predictions, not affect error and disturbance.
The parameters are given by
\begin{equation}
A=\sigma_x, \quad B=\sigma_y,\quad
\ket{\psi}=\cos\frac{\theta_{\psi}}{2}\ket{+z} +
e^{i \phi_{\psi}}\sin\frac{\theta_{\psi}}{2}\ket{-z}
\end{equation}
leading to
\begin{eqnarray}
\label{eq:sigmaA_psi_var}
\sigma(A)\quad &=&\quad\sqrt{1-\cos^2\phi_{\psi}\sin^2\theta_{\psi}}
\\
\label{eq:sigmaB_psi_var}
\sigma(B)\quad&=&\quad \sqrt{1-\sin^2\phi_{\psi}\sin^2\theta_{\psi}}
\\
\label{eq:limit_psi_var}
\frac{1}{2}\bra{\psi}[A,B]\ket{\psi}&=&|\cos\theta_{\psi}|
\end{eqnarray}

\begin{figure*}
\begin{center}
\includegraphics[width=17cm,keepaspectratio=true]{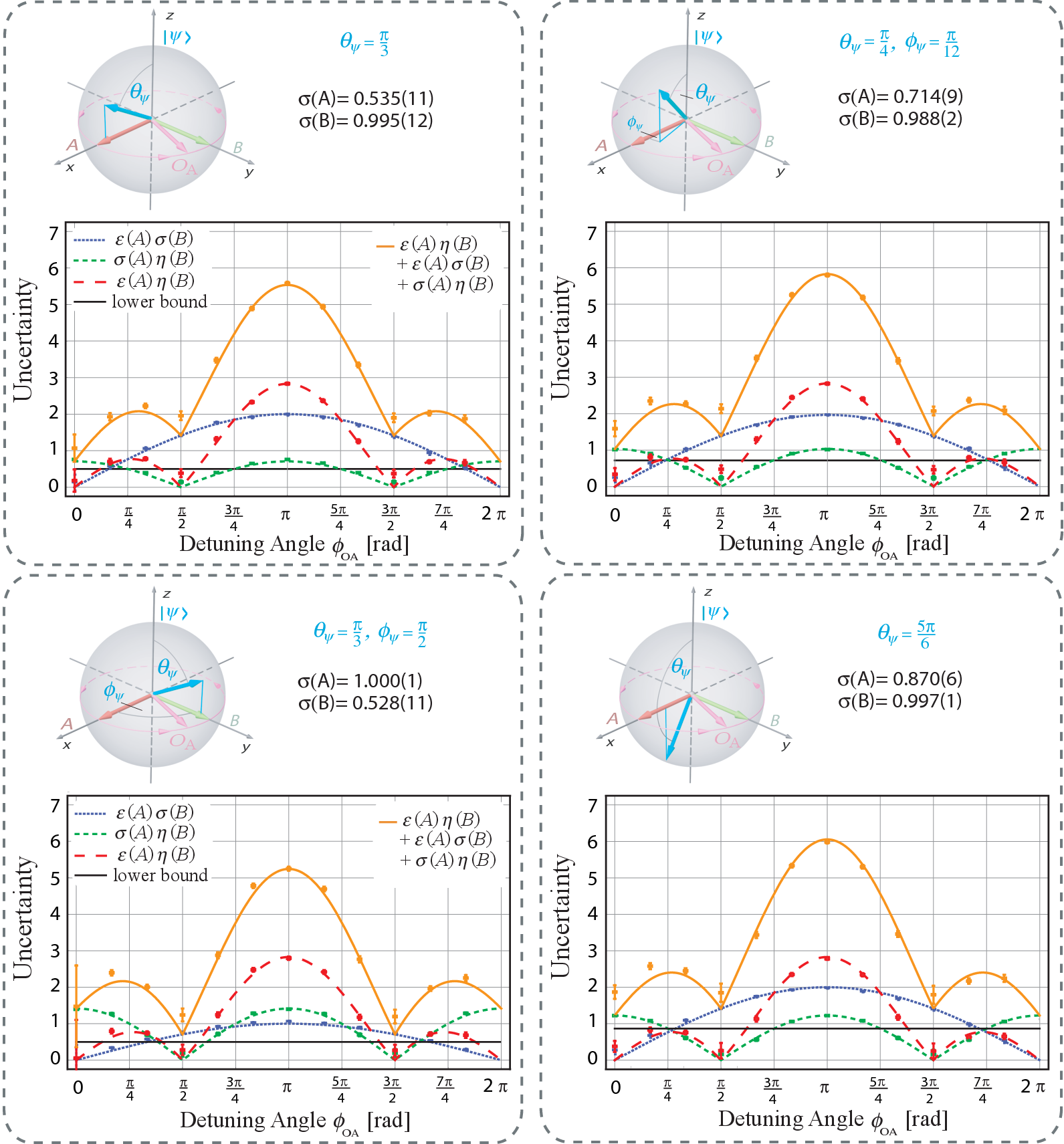}
\caption{(Color online) Variations of the initial state $\ket{\psi}$. The observables $A$ and $B$, the initial state $\ket{\psi}$ and the chosen path for the output operator $O_A$ are depicted on the Bloch sphere. The theoretically predicted values for $\sigma(A)$, $\sigma(B)$ and the lower limit are given by Eqs.\,(\ref{eq:sigmaA_psi_var}\,-\,\ref{eq:limit_psi_var}) and the theory curves for $\epsilon(A)$ and $\eta(B)$ by \Eq{error_dist_std_config}.}
\label{fig:psi_variations}
\end{center}
\end{figure*}

Since $O_A$ is varied along the equator no change for error and disturbance in comparison to the standard configuration is expected and their theory curves are thus given by \Eq{error_dist_std_config}. After dividing with the corresponding standard deviations this can be verified from the $\epsilon(A)\sigma(B)$- and $\sigma(A)\eta(B)$-curve in Fig.\,\ref{fig:psi_variations}.

The plots show that the product of error and disturbance is indeed independent from the initial state. However, the lower limit changes and the regions where Heisenberg's error-disturbance relation is fulfilled become larger when $\ket{\psi}$ approaches the $xy$-plane.

The left-hand-side of Ozawa's relation also contains the standard deviations and therefore changes with the initial state and still lies above the limit.

\section{Discussion}

In our previous letter \cite{Erhart12}, we reported an experiment for the particular case where error and disturbance of a neutron's spin-component measurements obey Ozawa's new relation \Eq{ozawa} but violate the old one \Eq{heisenberg_general} in the whole range of the experimental parameter. The old relation is extended from Heisenberg's original relation \Eq{Hei27} between the error of a position measurement and the induced disturbance on the momentum.
Although mainly motivated by thought experiments and in its mathematical derivation based on unjustified assumptions, Heisenberg's old relation for measurement error and disturbance prevailed and was regarded as a peculiarity of quantum mechanics for a long time.
The important step for the test of error-disturbance uncertainty relations is to give clear and consistent definitions of error and disturbance for quantum measurements. This is done by Eqs.(\ref{eq:error_def}) and (\ref{eq:dist_def}).
Here, we again investigate these relations in the $\frac{1}{2}$-spin system, but enlarge our parameter range in comparison to \cite{Erhart12}. For projective spin-$\frac{1}{2}$ measurements, error and disturbance are determined by three (Bloch) vectors $\vec{a}$, $\vec{b}$ and $\vec{o}_a$, characterizing the spin observables $A$ and $B$ and the actual measurement operator $O_A$. In particular, the error (disturbance) only depends on the deviation angle between  $\vec{a}$ ($\vec{b}$) and $\vec{o}_a$, which is described in \Eq{error_spin} (\Eq{dist_spin}) and depicted in Fig.\,\ref{fig:error_bloch} (Fig.\,\ref{fig:dist_bloch}). It is worth noting here that error and disturbance are independent from the initial quantum state which follows straight-forward from the properties of the Pauli-matrices.

Our measurement strategy is based on the expansion of error and disturbance
in terms of expectation values of outcomes in three different
input states given by Eqs.\,\eq{error_5terms} and \eq{dist_5terms} reminding of a quantum process tomography \cite{NielsenChuang_JModOpt_1997, Poyatos}.
There is another experimental strategy based on weak values \cite{lund_wiseman_njop2010, Steinberg_PRL_2012, Hofmann_arxiv_2012}.
They insist that ``if the system is weakly measured before the measurement apparatus the precision (error) and disturbance can be \emph{directly} observed in the resulting weak values'' \cite{Steinberg_PRL_2012} and claim our method to be indirect.
Although it is not clear in what sense the measurements are direct and indirect, we consider here possible arguments:
The determination of error and disturbance is done
(i) whether by using a single incident state or a combination of some states  and/or
(ii) whether for single events or for an ensemble.
For the point (i), we point out the fact that the incident state is affected by some operation before the measurements A and B in both cases.
In the weak-value strategy, weak measurements actually demands interaction with the system, which inevitably makes backreaction on the system (however weak they may be), otherwise no information of the system will be derived: after interactions through needed weak-value measurements, the incident state is no more the same as it was.
In contrast,  the measurements of A and B are done just after the incident state goes through the operation of $\1$, A, B, A+$\1$, or B+$\1$ in the tomographic strategy. Only the difference between them is that the operation is weak or strong.
For the point (ii), as is stated in one of the above papers, ``Experimentally, the problem with weak values is that they cannot be confirmed by precise measurements on individual systems'' \cite{Hofmann_arxiv_2012}: experimental determination of the error and disturbance requires measurements on a set of an ensemble, which is a common feature of both approaches. This is a clear statement of abandonment of determining the error and disturbance for single events by using neither strategies.
Another point is to be mentioned:
%that accuracy of determination by both strategies should be mentioned: 
in comparison of the two graphs in Fig.\,5 of \cite{Erhart12} (respectively, the corresponding part of Fig.\,\ref{fig:standard_config} in this work) and Fig.\,4 \cite{Steinberg_PRL_2012}, the accuracy of the former is clearly much higher than the latter.

Conscientious readers may presume without difficulty that the new uncertainty relation could be reformulated as 
\begin{equation}
(\epsilon(A)+\sigma(A))\cdot(\eta(B)+\sigma(B)) \ge |\bra{\psi}[ A , B]\ket{\psi}|
\end{equation}
by just adding Robertson's and Ozawa's inequality  (Eqs.\,\eq{robertson} and \eq{ozawa}).
Although it seems appealing to consider the left hand side as the product of a newly defined ``overall'' error and ``overall'' disturbance,
given by the sum of measurement-induced error for $A$ 
[disturbance on $B$] and measurement-independent 
intrinsic fluctuation
(standard deviation) of $A$ [$B$], the ``overall'' lower bound is less tight  
than Eqs.\,\eq{robertson} and \eq{ozawa} considered separately.

Finally, we are aware that a completely different quantification of uncertainty relations, viz, in terms of entropy, has been developed. For instance, a relation for position and momentum was derived first \cite{Mycielski1975}, followed by a generalization for arbitrary pairs of observables \cite{Deutsch1983}. An improved version suggested in \cite{Kraus_PhysRevD_1987} and proven in \cite{Maassen1988} has the form
\begin{equation}
\label{eq:EntropyUR}
H(A) \cdot H(B) \ge - \log{2} \max_{j,k}{\|\langle a_j|b_k\rangle\|}
\end{equation}
where $H(A)$ [$H(B)$] denotes the Shannon entropy of the probability distribution of the outcome and $|a_j\rangle$ [$|b_k\rangle$] represent non-degenerating eigenstates of A [B].
Recently, a stronger entropic uncertainty relation for an entangled system was presented \cite{Berta2010} and studied in an optical setup \cite{Li2011}. The entropic uncertainty relations describe informational contents and do not refer to the interaction of quantum measurements. They rather represent a generalization of Robertson's relation (\Eq{robertson}) for probability distributions for which the standard deviation has little physical meaning. Though, it would be interesting if the universally valid uncertainty relation of Ozawa (\Eq{ozawa}) also has an entropic counterpart.

\section{Conclusion}

In Heisenberg's original formulation, the uncertainty principle refers to a relation 
for the error in the measurement of a certain observable and the thereby induced 
disturbance on another observable. 
However, his famous relation has been based on thought experiments or
a rather heuristic argument with unsupported assumptions.
A recent development of rigorous treatments of quantum measurement has 
enabled us to reformulate the error-disturbance relation and lead to a universally 
valid relation. 

In the experiment presented here, we have investigated this relation 
for projective neutron spin measurements. The two non-commuting observables under consideration are spins along different directions which are measured successively. By using a tomographic procedure, based on applying three different states generated from the initial system state on the joint measurement apparatuses, the error of the first measurement and the disturbance induced on the second observable can be determined from the output intensities. By detuning the measurement axis of the first apparatus we can study the variation of error and disturbance. Together with the Bloch vector of the initial system state, the directions of the two spin observables and of the measurement axis constitute the relevant parameters determining all quantities occurring in the uncertainty relations. 

In our neutron-optical experimental runs, we have at first fixed the observables and the initial state, and then varied the detuned measurement axis along equilatitude circles on the Bloch sphere. For various representative configurations, the measured values for error and disturbance are in excellent agreement with the theory curves. 
They are independent of the initial state and are solely determined by the enclosed angle between the measurement axis and the first or second spin axis, respectively. 
The results demonstrate that Heisenberg's error-disturbance relation can always be violated for a non-vanishing lower limit.
In contrast, Ozawa's new inequality always holds.
Furthermore, we conclude that increasing error does not always lead to decreasing disturbance and vice versa in spin measurements. 
Such a reciprocal behavior occurs only in certain areas and along certain directions. 
Thus, our results give an experimental demonstration that the generalized form of Heisenberg's error-disturbance relation has to be abandoned.

\end{document}